\documentclass[final,3p,times,twocolumn]{elsarticle}

\usepackage{amssymb}

\usepackage{xcolor} 
\usepackage{mathtools, cuted} 

\usepackage[final]{changes}
\definechangesauthor[color=red]{MAN}
\setauthormarkuptext{} 

\journal{Chaos, Solitons \& Fractals}

\begin{document}

\begin{frontmatter}



\title{Universal 
Scaling of 
Electron Transmission for Nearly Ballistic 
and Quantum Dragon Nanodevices}


\author[inst1]{Mark A.\ Novotny}

\affiliation[inst1]{organization={Department of Physics and Astronomy, and HPC$^2$ Center for Computational Sciences},
            addressline={ Mississippi State University}, 
            city={Mississippi State},
            postcode={39762-5167}, 
            state={MS},
            country={USA}}

\author[inst2]{Tom{\'a}{\v s} Novotn{\'y}} 

\affiliation[inst2]{organization={Department of Condensed Matter Physics, Faculty of Mathematics and Physics, Charles University},
            addressline={Ke Karlovu 5}, 
            city={Prague},
            postcode={CZ-121 16}, 
            country={Czech Republic}}

\begin{abstract}
We predict two different universal scaling regimes 
for the quantum transmission of 
metallic nanodevices following the 
addition of a small amount of uncorrelated disorder.  
A nanodevice is connected to two thin semi-infinite uniform leads, 
and the Non-Equilibrium Green's Function (NEGF) methodology 
yields the electron transmission ${\cal T}(E)$ 
as a function of the injected electron energy $E$.  
Ballistic nanodevices have no disorder and 
have ${\cal T}(E)$$=$$1$ for all 
$E$ that allow electron propagation in the leads.  
Quantum dragon nanodevices can have extremely strong  
properly correlated disorder, 
and still have ${\cal T}(E)$$=$$1$ for all $E$.  
Additional uncorrelated site disorder leads to 
Fano resonances in ${\cal T}(E)$.  
Averaging over the uncorrelated disorder we predict 
\added[id=MAN
]{using perturbation theory }
two universal scaling regimes for 
${\cal T}_{\rm ave}(E)$.  
\added[id=MAN
]{The functional form of both universal scaling regimes depend on the 
device length and width, energy, and variance of the 
uncorrelated disorder.  The second scaling regime,
valid for small but somewhat larger uncorrelated disorder than 
the first scaling regime, 
also has the form dependent on the density of states of the system.}
These two scaling regimes 
are demonstrated \added[id=MAN]{to be valid 
via 
large scale computer calculations.}
\end{abstract}




\end{frontmatter}






\section{Introduction}
\label{sec:1:Intro}




Disorder is disruptive to coherent electron transport, as it causes 
electron scattering 
\added[id=MAN]{\cite{DATTA1995}}. When a nanodevice is connected to two leads, the 
relevant quantity to measure is the transmission probability ${\cal T}(E)$ of 
electrons of energy $E$ that are injected into the incoming lead and are transmitted through the outgoing lead \added[id=MAN]{\cite{DATTA1995}}.  

Since the 1957 work of Landauer \cite{Landauer1957}, and generalizations \cite{BUT1985}, it has 
been known that the coherent electron transmission ${\cal T}(E)$ 
through a quantum nanodevice gives the 
electrical resistance of the nanodevice.  Hence the important quantity to calculate in 
any model Hamiltonian, and to consider in device construction, is the electron transmission 
${\cal T}(E)$ as a function of the energy $E$ of the injected electrons.  
\added[id=MAN]{
An excellent review of one framework for understanding quantum 
transport in mesoscopic systems is by using 
random-matrix theory, wherein the work bridges 
mesoscopic physics and statistical approaches initially developed for nuclear physics \cite{Beenakker1997}.  
}

Ballistic electron propagation may occur in perfectly ordered nanodevices, 
and has ${\cal T}(E)=1$ for a range of energies \cite{Karp2017}.  
Ballistic electron propagation can be used to create ballistic FETs (Field Effect 
Transistors) \cite{Javey2003}.  
More recently the possibility of forming qubits and quantum gates using ballistic nanodevices has been 
proposed \cite{Guo2009,Dragoman2016,Bertoni2000}.  
For quantum computing nanodevices the 
quality of the qubit and quantum gates increases as ${\cal T}$ approaches unity.  
Hence understanding systems that have an average transmission 
${\cal T}_{\rm ave}(E)\approx 1$ 
is of technological importance.  
Here the average is over small amounts of uncorrelated disorder.  
In 2007 Markussen {\it et al\/} \cite{MAR2007} studied nearly ballistic 
transport in silicon nanowires and concluded \lq the sample-to-sample
fluctuations depend on energy but not on doping density, 
thereby displaying a degree of universality'.  They used the 
concept of an average mean-free-path of the electrons, $L_e$, 
which is a 
well defined quantity for 
\added[id=MAN]{translationally invariant}
nanosystems with only a small number 
of \added[id=MAN]{added}
impurities.  If the length of the nanosystem is $L$ along the 
direction of electron flow, the nearly ballistic (small disorder) 
regime occurs when $L_e\gtrsim L$.  

\begin{center}
\begin{figure*}[tb]
\includegraphics[width=0.87\textwidth]{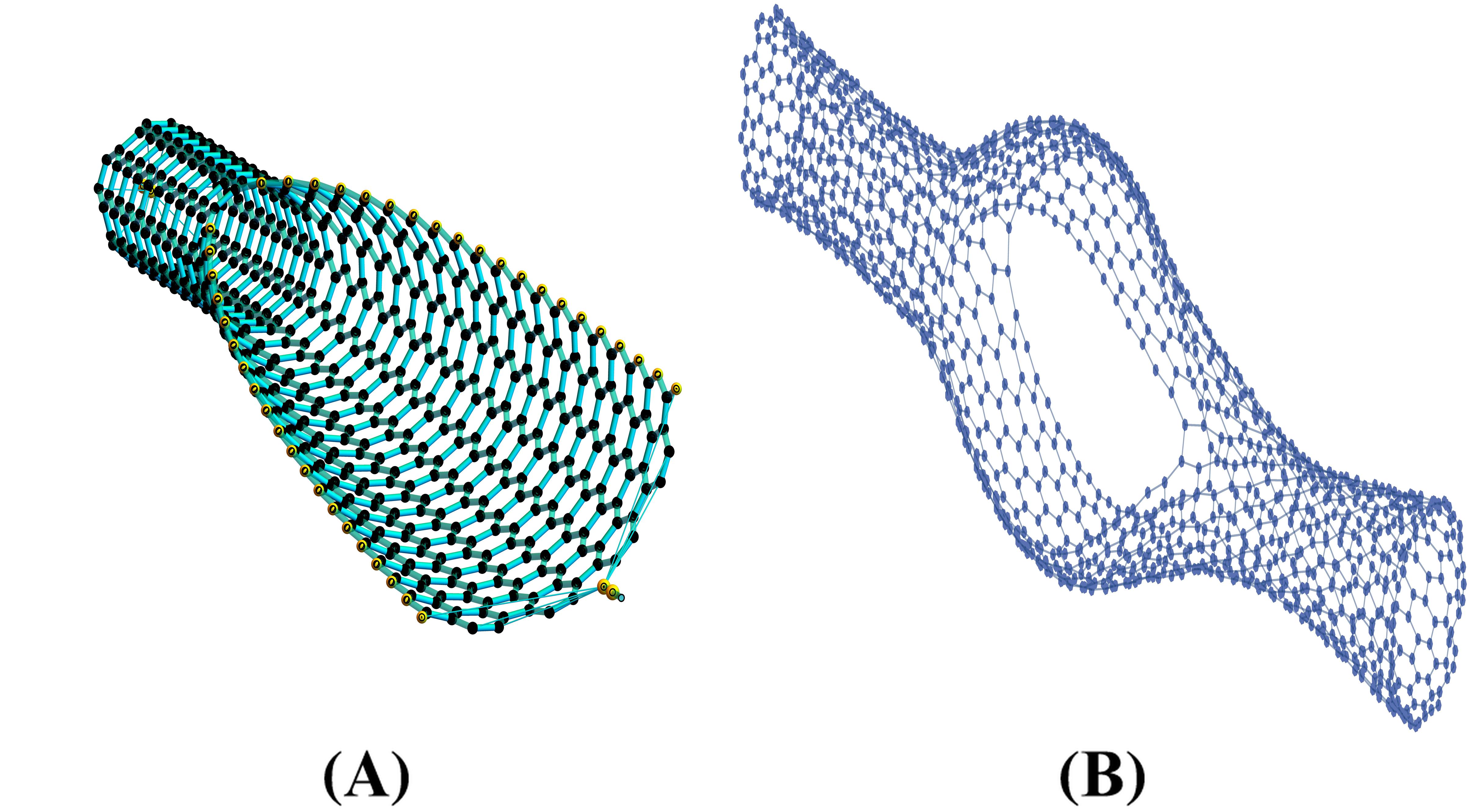}
\caption{
\label{Fig:SplitTubes}
\added[id=MAN
]{
Two quantum nanosystems
based on split armchair single-walled carbon nanotubes with 
$m$ atoms in each slice.  
With appropriate boundary conditions and input/output leads, 
both are quantum dragons with 
unit transmission, as in Eq.~(\ref{Eq:Landauer}), based on 
our model and method.  
(A) A nanodevice with $m$$=$$10$.  Compare directly with 
the graphical abstract of the experimental article \cite{He2023} 
of a sub-5~nm graphene-nanoribbon/carbon-nanotube which also has 
$m$$=$$10$.  
The black spheres represent a constant on site energy 
in a tight-binding model, with all on site energies the same.  
The yellow spheres, located on the boundary atoms and the atoms of the 
attached thin leads, have zero on site energy.  
The cyan cylinders represent the hopping energies, which are all 
the same value in the device and leads.  The radius of the 
cyan cylinders for the nanosystem-lead connections are proportional to the 
hopping strength.  
(B) Shows a double-cut armchair nanotube, without showing the two attached leads.  
To see the underlying structure of the graph, the Mathematica \cite{Mathematica} 
command {\tt GraphPlot} has been used.  
}
}
\end{figure*}
\end{center}

For 2D and 3D nanosystems, in the 
diffusive regime where $L_e\ll L$ 
there are
analytical predictions of the Dorokhov-Mello-Pereyra-Kumar theory 
\cite{Dorokhov1982,Mello1988} for the electrical conductance, wherein 
${\cal T}_{\rm ave}(E)$ is small.  For strong disorder 
$L_e$ may be ill defined, in which case in reference \cite{MAR2007} 
was defined an energy dependent 
length scale 
\begin{equation}
\label{Eq:L-T}
L_{\cal T}(E)=L\> {\cal T}(E)/\big(1-{\cal T}(E)\big)
\>.  
\end{equation}
Note for nearly ballistic nanodevices one has $L_{\cal T}(E)\approx L_e$ 
\cite{MAR2007}.

For even stronger disorder one has the localized regime where 
$\xi_{\rm A}\ll L$, and ${\cal T}_{\rm ave}(E)\ll1$.  
The seminal work initiated by Anderson 
\cite{ANDE1958,LAGE2009} 
showed in 1D (1 dimensional) materials models with short-range interactions 
that any amount of uncorrelated randomness gives that the wavefunction 
is exponentially localized, with a localization length $\xi_{\rm A}$. 
A nontrivial crossover from the diffusive regime to the Anderson localization 
regime \cite{Lopez2018PRB} is present.  
Notably, even in the 1D case long-range correlations in the disorder 
or long-range interactions in the Hamiltonian may 
produce extended states \cite{IZRA1999,IZRA2012,YU2015,CHOI2017}.  
The existence of long-range correlations in the disorder, 
and the interplay between diagonal and off diagonal disorder \cite{FLOR1989}, 
can lead to effective mobility edges and 
necklace states \cite{BERT2005} which occur at particular energies  
and have ${\cal T}(E_{\rm necklace})\approx 1$.  
A localized state would have ${\cal T}(E)\ll 1$ 
for all, or for finite $L$ all except a 
subset of measure zero, 
energies $E$.  

\added[id=MAN
]{
Usually ballistic transport refers to the regime in which charge carriers, such as electrons, propagate through a material without scattering from impurities, phonons, or other electrons. 
Hence such devices are uniform, with no disorder, and their translational invariance 
allows one to consider Bloch states for the wavefunctions and associated band structure analysis 
\added[id=MAN]{\cite{Buerkle2023}}   
The ballistic ($L_e\ll L_{\cal T}$), or nearly ballistic regime 
($L_e\gtrsim L_{\cal T}$), becomes particularly significant at the nanoscale, when device dimensions approach or fall below the electron mean free path. 
Carbon nanotubes (CNTs), due to their quasi-one-dimensional structure and high crystal quality, exhibit nearly ballistic transport over micron-scale lengths at room temperature, enabling the development of high-performance field-effect transistors 
\cite{White1998CNTballistic,yao2000high,javey2003ballistic,Franklin2022} 
as well as current densities up to $10^9$~A/cm$^2$. 
Similarly, graphene nanoribbons (GNRs) can support ballistic transport when fabricated with atomically smooth edges, and hence  can be 
made into FETs 
\cite{li2008chemically,lin2008graphene,Rizzo2019,Rizzo2020graphene}. The ability to achieve ballistic conduction in these low-dimensional systems is central to future nanoelectronic technologies that demand low power consumption and high switching speeds \cite{avouris2007carbon,PENG2014cntElectronics}.  
Researchers have also been able to synthesize nanotube-ribbon carbon 
nanosystems \cite{He2023}  (compare our Fig.~\ref{Fig:SplitTubes}A 
directly with the Fig.~1 of the experimental paper \cite{He2023}).  
Earlier work on partially unzipped CNTs illustrated other methods 
of synthesizing such systems 
\cite{Huang2009,Rangel2009}, giving further justification for the 
quantum dragon nanodevices in Fig.~\ref{Fig:SplitTubes} to be 
experimentally realizable.  
What has not previously been appreciated is the carbon nanosystems 
of  Fig.~\ref{Fig:SplitTubes} can have complete transmission 
even though they have disorder \cite{NOVO2014}.  With appropriate 
boundary conditions and properly connected leads, both 
nanodevices in Fig.~\ref{Fig:SplitTubes} are quantum dragon nanodevices as 
they exhibit unit transmission even though they have disorder due to the 
partial unzipping of the nanotube.  With a small amount of additional uncorrelated 
disorder these experimentally realizable carbon nanodevices \cite{He2023} 
are nearly quantum dragon nanodevices.  Therefore, it is important for future nanoelectronics to 
understand the transmission of nearly quantum dragon nanodevices, which is the 
emphasis of this article.  
}

In 2D, 3D, \added[id=MAN]{and higher D} 
the existence in some energy range of 
extended (not localized) 
states for linear operators with extensive disorder is an open problem in 
mathematical physics 
\added[id=MAN]{\cite{Stolz2011}}.  
The scaling theory for typical disorder 
\added[id=MAN]{\cite{ABRA1979,Lee1985}}
for D$\ge$$2$ suggests in 3D that for 
small enough uncorrelated typical disorder 
extended states almost surely occur, but for strong enough disorder 
the typical case is for only localized states to exist.  
2D is a marginal dimension, as shown in \cite{ABRA1979}
for the typical case there is no true metallic 
behavior, with decreasing ${\cal T}(E)$ exhibiting 
a crossover from logarithmic to exponential in the device length $L$.  

In 2012 Rodriguez, Chakrabarti, and R{\"o}mer showed that 
in D$>$1 it is possible with global correlation of disorder to 
engineer extended states in highly disordered systems \cite{2012PRBdragonRodriguez}.  
In 2014 one of the authors showed that for electron transport 
in models of nanodevices disorder needed to only be correlated 
in slices perpendicular to the direction of electron flow to allow extended states \cite{NOVO2014}.  
Such nanodevices have ${\cal T}(E)$$=$$1$ for a wide range of energies, 
and were named quantum dragon nanodevices \cite{NOVO2014,Inkoom_2018}.  
Here we demonstrate that quantum dragon nanodevices need only 
have disorder correlated {\it locally\/}, as opposed to correlated 
with any size of the nanodevice.  We have previously shown 
that quantum dragon nanodevices may show order-amidst-disorder 
\cite{Novotny_2021,Novotny_2023} where the strongly disordered 
quantum dragon nanodevice has order in commonly measured electron transport quantities such as the local density of states (LDOS).  

In this paper we study only {\it local\/} disorder correlations 
(atypical disorder), with an emphasis on the interplay of diagonal and 
off diagonal disorder, in 2D and 3D (and 2D+3D) models 
of ordered ballistic nanodevices and quantum dragon nanodevices.   
In particular, we are interested in how ${\cal T}_{\rm ave}(E)$ 
depends on $E$, the strength of random added on site disorder, 
as well as the lengths $L$ and $L_{\cal T}$.  In particular, we 
start with a nanodevice model that has 
${\cal T}(E)$$=$$1$ for a range of 
energies, and add a small amount of uncorrelated random on site 
disorder.  
The transmission average ${\cal T}_{\rm ave}(E)$ is an average  
only over different samples of this added uncorrelated random disorder. 
We will demonstrate two different {\it universal\/} scaling 
regimes for ${\cal T}_{\rm ave}(E)$.

We study electron transport and obtain the electrical 
conductance $G$ (the electrical resistance is
$R=1/G$) of a model for a nanodevice.  
Uniform semi-infinite 1D thin leads are attached to the nanodevice.    
The leads have a hopping strength which we take to 
be our unit of energy, and have zero on site energy 
(defining our zero of energy).  
Therefore, electrons with energies in the range $-2$$<$$E$$<$$2$ can propagate in 
the leads (see \ref{CSaF_AppA}), and have a dispersion relation 
$\cos\left(q_{\rm lead} a\right)=-E/2$ where 
$a$ is the lattice spacing between atoms in the leads and 
$q_{\rm lead}$ is the wavevector of the electrons in the lead.  
When the leads are attached to the device, and connected at infinity to macroscopic 
reservoirs with slightly different chemical potentials, an electrical current 
flows through the device.  
We take the current flow in the device as being from slice $j$ to $j$$+$$1$.  
The zero temperature, low bias electrical conductance $G$ is related to the 
electron transmission ${\cal T}(E)$ at the Fermi energy $E_{\rm F}$ through the 
Landauer relations
\cite{Landauer1957,BUT1985} 
\begin{equation}
\label{Eq:Landauer}
G  \! = \!  
\left\{
\begin{array}{lclcl}
\! G_0\> {\cal T}\left(E_{\rm F}\right) 
& \stackrel{{\cal T}(E)\rightarrow 1}\longrightarrow 
& G_0  
& \>
& {\rm two \> probe} \\
\\
\! G_0\>\frac{{\cal T}\left(E_{\rm F}\right)}{1-{\cal T}\left(E_{\rm F}\right)} 
& \stackrel{{\cal T}(E)\rightarrow 1}\longrightarrow  
& \infty 
&  & {\rm four \> probe} 
\end{array}
\right.
\end{equation}
as seen also experimentally \cite{dePicciotto2001}.  
The Landauer analysis can be extended to finite voltage bias
\cite{Bagwell1989}, but in this paper we focus only on low bias.

This paper is organized into six additional sections, supported by five appendices.  
Section~\ref{sec:2:ModelMethod} describes the tight binding model, the 
NEGF (Non-Equilibrium Green's Function) method to calculate ${\cal T}(E)$, 
and the construction of quantum dragon nanodevices.
Section~\ref{sec:3:Fano} provides background on Fano resonances that 
are important to theoretically calculate ${\cal T}_{\rm ave}(E)$.  
An overview of the general dependence of ${\cal T}_{\rm ave}(E)$ on 
random uncorrelated disorder of strength $\delta$ is 
presented in Sec.~\ref{sec:4:NearQuantumDragon}.
The universal scaling regime for ${\cal T}_{\rm ave}(E)$ for  
very small $\delta$ 
is presented in Sec.~\ref{sec:5:SmallDelta}.  
Section~\ref{sec:6:ScaleDOS} demonstrates the second universal 
scaling regime, valid for larger $\delta$.
Finally Sec.~\ref{sec:7:ConcDisc} presents 
conclusions and discussion of our results.  
\added[id=MAN]{
Although quantum dragon nanodevices have been 
discussed previously 
\cite{NOVO2014,Novotny_2021,Novotny_2023}, 
this article is the first to quantitatively predict and test 
universal scaling for nearly quantum dragon nanodevices.  
The predicted scaling also is valid for small disorder 
in any nanodevice exhibiting ballistic electron transport 
with a Hamiltonian approximated by a tight binding model, 
including ballistic electron transport 
in carbon nanotubes and and graphitic nanoribbons.  
}

\section{Model and Method}
\label{sec:2:ModelMethod}

\subsection{Tight Binding Model}
We study the standard single-orbital tight binding model.  
This is sometimes also called the Anderson model.  
We examine the model on a graph 
\cite{CHAR2012} of $N$ vertices with edges which can be 
viewed as being in 2D, 3D, or 2D+3D.  
We assume the graph vertices can be partitioned 
into $\ell$ slices each with $m$ vertices.  
Index the vertices within a 
slice by $i=1,2,\cdots, m$ and the slices by $j=1,2,\cdots,\ell$.  
We here restrict ourselves only to graphs with intra-slice edges 
(vertices with the same value of $j$) and inter-slice edges 
confined to being between vertices in slice $j$ and in 
adjacent slices $j\pm1$. 
As a model for materials, the graph vertices represent 
the locations of  
atoms, and the graph edges are hopping paths, 
sometimes called bonds, for electrons between the atoms due to the electron wavefunction overlap of 
the two atoms labeled $i,j$ and $i',j'$.  
We study only graphs which may represent actual nanomaterials,
by limiting ourselves to $D\le 3$.  We further 
restrict only to very short-range bonds, 
either nearest neighbor (nn) 
or next-nearest neighbor (nnn) bonds.  
The Hamiltonian 
on the graph has the form 
\begin{equation}
\label{Eq:HamTB}
\begin{array}{lcl}
{\cal H} & \> = \> & 
\sum\limits_{j=1}^\ell \sum\limits_{i=1}^m \> \epsilon_{i,j} \>\> 
{\hat c}_{i,j}^\dagger {\hat c}_{i,j}
\\
& & \> - \>
\sum\limits_{\left\langle i,j;i',j'\right\rangle}
t_{i,j;i',j'} \left(
{\hat c}_{i,j}^\dagger {\hat c}_{i',j'}
+ {\hat c}_{i',j'}^\dagger {\hat c}_{i,j}
\right)
\>.
\end{array}
\end{equation}
The graph has $N=\ell m$ vertices ($N$ atoms in the device), 
and each has an associated on site energy $\epsilon_{i,j}$.  
The creation (annihilation) operators 
are ${\hat c}_{i,j}^\dagger$ (${\hat c}_{i,j}$).  
Every bond present has a hopping term of strength $t_{i,j;i',j'}$, 
and the second sum is over all edges of the graph.  
Note the restriction to nn and nnn bonds is not necessary \cite{NOVO2014} to find quantum 
dragons, but gives more physical quantum dragon nanodevices.

For one type of our universal scaling we will need the 
Density of States DOS$(E)$ of the nanodevice.  
We calculate DOS$(E)$ in the normal 
box-counting fashion.  Namely the Hamiltonian 
${\cal H}$ of Eq.~(\ref{Eq:HamTB}) is 
diagonalized yielding $N$ eigenvalues. The DOS$(E)$ is 
obtained by a box-counting method, namely counting the number of eigenvalues in a 
given interval $\Delta E$ that fall within the interval 
$\left[E-\frac{1}{2}\Delta E,E+\frac{1}{2}\Delta E\right]$.  
The DOS is calculated for the device only, without 
any attached leads.  

The tight binding model we study in Eq.~(\ref{Eq:HamTB}) is the 
traditional model to study 
conductance through nanodevices.  Although the model is a 
low-level approximation for actual materials, it nevertheless has been 
well studied and applied to understanding properties of materials. 
For example, with only nn hopping terms the tight binding 
model has been used to study $\pi$-orbital graphene nanoribbons 
\cite{NAKA1996}, but of course for direct comparison to a material 
an {\it ab initio\/} method is 
preferable \cite{SON2006}.  
In general the hopping terms may be complex numbers, 
but in this article we take the hopping terms to be a real number and hence 
study only 
zero external magnetic fields. 

\begin{center}
\begin{figure}[tb]
\includegraphics[width=0.47\textwidth]%
{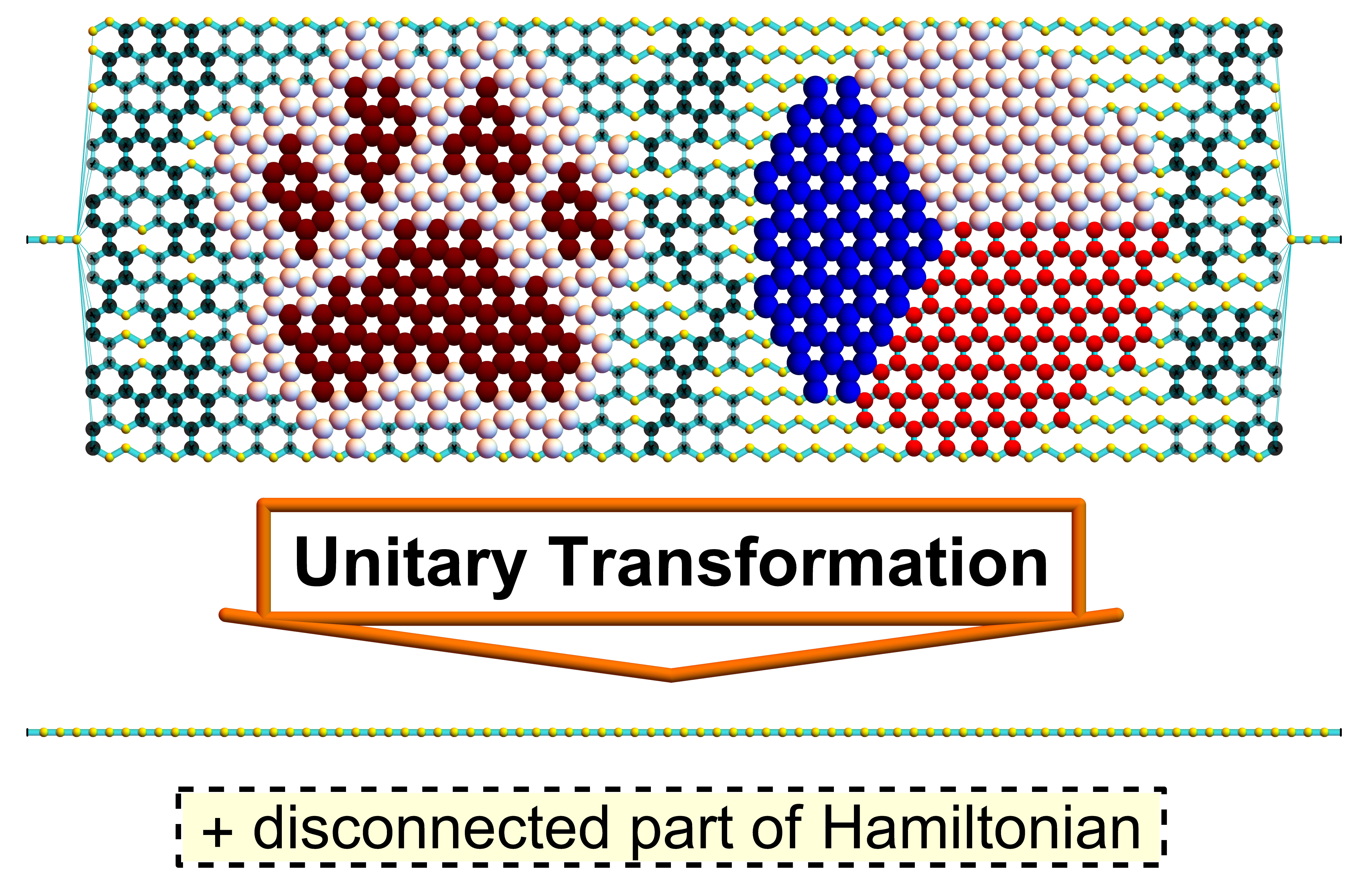}
\caption{
\label{Fig:FindDragon}
Technique to show complete electron transmission for atypical 
strongly disordered nanosystems.  
See text in Sec.~2.3.2 for full description 
of this quantum dragon nanodevice\added[id=MAN]{, including the 
color coding.  
The top diagram is the quantum dragon nanodevice with 
$m$$=$$16$ and $\ell$$=$$73$ depicted in real space.  
In the basis after the unitary transformation, depicted in the bottom,
the system is a uniform wire plus a disconnected part of the Hamiltonian.  
A unitary transformation does not affect the transport properties, hence 
the device is a quantum dragon with ${\cal T}(E)$$=$$1$ for 
all $-2$$<$$E$$<2$.
}  
}
\end{figure}
\end{center}

\subsection{NEGF (Non-Equilibrium Green's Function) Method}
We utilize the standard NEGF method
\cite{DATTA1995,TODO2002,FERR2012,HIRO2014}.  
\ref{CSaF_AppA} gives a NEGF analysis of thin, uniform wires 
within the tight binding formalism.  
We connect two such semi-infinite wires to our nanodevice, 
as in Fig.~\ref{Fig:FindDragon}.  
The hopping terms between the incoming wire and the 
left-most slice of the device is given by a length $m\ell$ vector 
$| L\rangle$.  Since only the first $m$ device atoms are connected to 
the incoming wire only the first $m$ elements of $| L\rangle$ 
are non-zero.  
Similarly, the hopping terms between the outgoing wire and the 
right-most slice of the device is given by a length $m\ell$ vector 
$|R\rangle$, with all elements zero except for the last $m$ 
which are the hopping terms between the $m$ atoms in the 
right-most slice of the device.  
As in \ref{CSaF_AppA} with ($a=1$, $\epsilon_{\>\rm Lead}=0$, 
and $t_{\>\rm Lead}=1$) the wavevector in the lead is given by 
\begin{equation}
\label{Eq:NEGF:01}
\cos\left(q_{\rm Lead}\right) = -\>\frac{E}{2} 
\>.
\end{equation}
Define the quantity
\begin{equation}
\label{Eq:NEGF:02}
\xi(E) \> = \> e^{-i q_{\rm Lead}} 
\> = \> 
-\> \frac{E}{2} \> - \>i \> \frac{\sqrt{4-E^2}}{2}
\>.
\end{equation}
The self energies for the left (L) and right (R) leads are
\begin{equation}
\label{Eq:NEGF:03}
{\bf\Sigma}_L(E)  =  -\xi^* |L\rangle \langle L| 
\quad {\rm and} \quad
{\bf\Sigma}_R(E)  =  -\xi^* |R\rangle\langle R|
\>,
\end{equation}
respectively.  
The retarded Green's function for the device with the 
two leads is 
\begin{equation}
\label{Eq:NEGF:04}
{\cal G}(E)=\Big(\>E{\bf I}_N-{\cal H}-{\bf \Sigma}_L(E)-{\bf \Sigma}_R(E) \> \Big)^{-1}
\>.
\end{equation}
\added[id=MAN]{Here ${\cal H}$ is the device Hamiltonian of 
Eq.~(\ref{Eq:HamTB}) and 
${\bf I}_N$ is the $N$$\times$$N$ identity matrix.}

The local density of states, LDOS$(E)$, for the propagating electrons 
is given by the diagonal elements of ${\cal G}(E)$, namely
\begin{equation}
\label{Eq:NEGF:LDOS}
{\rm LDOS}(E) \> = \> 
- \frac{1}{\pi} \>{\cal I}m\Big({\cal G}_{i,j;i,j}(E)\Big)
\>.
\end{equation}
The LDOS$(E)$ is an experimentally measurable quantity in nanosystems 
\cite{LDOS2005exp,LDOS2023exp}.  
Note the LDOS is calculated once the leads are 
attached to the device, and hence 
contains a projection 
operator onto only the propagating electrons.  
In contrast, in the new basis after the similarity transformation 
depicted in Fig.~\ref{Fig:FindDragon}, the 
DOS is calculated using both the nanodevice wire portion and the 
disconnected part as in Fig.~\ref{Fig:FindDragon}, since 
the DOS is calculated in the basis where the Hamiltonian is diagonal.  
In terms of the mathematics, 
the LDOS is calculated from the Green's function 
of Eq.~(\ref{Eq:NEGF:LDOS})
while the DOS is calculated from the Hamiltonian 
as described in Sec.~2.1.  
This means for a quantum dragon the LDOS does not contain 
any contribution from the disconnected \lq pieces', 
as in 
Fig.~\ref{Fig:FindDragon}.  

Introduce the broadening functions for the two leads 
\begin{equation}
\label{Eq:NEGF:05}
\begin{array}{lclcl}
{\bf\Gamma}_L(E) 
& \> = \> & i\left({\bf\Sigma}_L-{\bf\Sigma}_L^\dagger\right) 
& \> = \> & 
\gamma(E)|L\rangle\langle L|
\\
&{\rm and}& \\
{\bf\Gamma}_R(E) & = & 
i\left({\bf\Sigma}_R-{\bf\Sigma}_R^\dagger\right) 
& = & \gamma(E)|R\rangle\langle R|
\end{array}
\end{equation}
with the definition $\gamma(E)\equiv\sqrt{4-E^{2}}$. 
It is important to note that $|L\rangle$ and $|R\rangle$ 
are not wavefunctions, but rather are vectors containing hopping terms 
of the Hamiltonian between the leads and the ends of the device.  
Thus finally we obtain for the electron transmission 
probability the 
\added[id=MAN]{NEGF} 
expression \cite{DATTA1995,TODO2002,FERR2012,HIRO2014} 
\begin{equation}
\label{Eq:NEGF:06}
{\cal T}(E) 
\> = \>
{\rm Tr}\Big(\> {\bf \Gamma}_L(E) \> 
{\cal G}(E) \> {\bf \Gamma}_R(E) \> 
{\cal G}^\dagger(E) \> \Big)
\>.
\end{equation}

\subsection{Quantum Dragon Nanodevices}
A quantum dragon nanodevice \cite{NOVO2014}
is one with transmission ${\cal T}(E)$$=$$1$ 
for a wide range of $E$, while the device has very strong disorder 
\cite{Inkoom_2018,Novotny_2021,Novotny_2023}.  
Disorder is disruptive to coherent 
electron transmission so it was expected for 
almost all $E$ one would have ${\cal T}(E)$$\ll$$1$, 
hence the existence of quantum dragons was unexpected.  
Even knowing quantum dragon devices exist, the usual theoretical 
tools of Bloch wavefunctions and band structure cannot be used.  Due to 
the very strong disorder, even the wavevector in the disordered device is ill defined.  
(The Fourier transform of the disordered model can be performed, 
but the lack of translational invariance makes a physical association with 
the wavevector of an electron quasi-particle problematic.)  
The way to find and analyze the atypical disordered cases 
which are quantum dragon nanodevices 
is sketched in Fig.~\ref{Fig:FindDragon}.  

\subsubsection{Finding dragons: General method}

The Hamiltonian in Eq.~(\ref{Eq:HamTB}) for an $m\times\ell$ graph is 
a $\ell m\times \ell m$ matrix ${\cal H}$.  We perform a 
similarity \added[id=MAN]{(here unitary)} transformation 
on ${\cal H}$, as depicted in Fig.~\ref{Fig:FindDragon}, 
in order to block diagonalize the Hamiltonian.  We study only 
atypical locally correlated disorder where this similarity transformation 
can be obtained.  We have restricted ourselves to 
Hamiltonians which have a block-tridiagonal structure, 
namely written for $\ell=6$ slices a form
\begin{equation}
\label{Eq:HamMat}
{\cal H} = 
\left(\begin{array}{cccccc}
{\bf A}_1 & {\bf B}_{1,2} & {\bf 0} & {\bf 0} & {\bf 0} & {\bf 0} \\
{\bf B}_{1,2}^\dagger &{\bf A}_2 & {\bf B}_{2,3} & {\bf 0} & {\bf 0} & {\bf 0} \\
{\bf 0} & {\bf B}_{2,3}^\dagger &{\bf A}_3 & {\bf B}_{3,4} & {\bf 0} & {\bf 0} \\
{\bf 0}  & {\bf 0} & {\bf B}_{3,4}^\dagger &{\bf A}_4 & {\bf B}_{4,5} & {\bf 0} \\
{\bf 0}  & {\bf 0} & {\bf 0} & {\bf B}_{4,5}^\dagger &{\bf A}_5 & {\bf B}_{5,6} \\
{\bf 0} & {\bf 0}  & {\bf 0} & {\bf 0} & {\bf B}_{5,6}^\dagger &{\bf A}_6 \\
\end{array}\right)
\end{equation}
where each submatrix is $m\times m$.  
We restrict our studied nanodevices so all 
$\ell$ intra-slice matrices ${\bf A}_j$ (which contain all 
on~site energies and all intra-slice hopping terms) 
as well as all ${\bf B}_{j,j+1}$ inter-slice matrices (which contain 
all inter-slice hopping terms) have a common eigenvector, ${\vec v}_{\rm Dragon}$.  
Note the ${\bf A}_j$ are Hermitian, while the ${\bf B}_{j,j+1}$ 
need not be Hermitian.  
In fact, the method has been generalized 
so the ${\bf B}_{j,j+1}$ are not even square matrices 
\cite{NOVO2014}.  
In this paper we will choose ${\vec v}_{\rm Dragon}$ to have 
every element equal to $1/\sqrt{m}$.  
Ref.~\cite{Novotny_2023} generalizes this choice for 
a unitary matrix.  

The effect of the similarity transformation is 
depicted in Fig.~\ref{Fig:FindDragon}.  
In the new basis one block of size $\ell\times\ell$ is the Hamiltonian 
of a uniform wire with 
$\ell$ sites, and only this part is connected to the external leads.  
Here the similarity transform consists of 
the product of two unitary matrices.  
One unitary matrix $\mathbf{X}_{N}$ consists of a block diagonal matrix 
with $\ell$ blocks each being a $m\times m$ discrete Fourier transform matrix, 
or Vandermonde matrix, with the $k_1,k_2$ element $\omega^{(k_1-1)(k_2-1)}/\sqrt{m}$ 
with the $m^{\rm th}$ root of unity $\omega=\exp(-i 2\pi/m)$.  The second unitary matrix 
$\mathbf{P}_{N}$ is a permutation matrix which puts the first element of the $k^{\rm th}$ $m\times m$ 
block into the $k^{\rm th}$ position, 
while shifting all other elements out of the 
first $\ell\times\ell$ matrix block. The resulting 
unitary similarity transformation matrix $\mathbf{U}_{N}$ has a specific structure.  
The rectangular block submatrix of $\mathbf{U}_{N}$ 
connecting the original site basis $\alpha\equiv\{i,j\}$ 
(with $i=1\dots m,\>\>j=1\dots \ell$)
with the uniform wire part (labelled by $\eta=1,\dots,\ell$) 
in the rotated basis is 
\begin{equation}\label{eq:similarity-matrix}
(\mathbf{U}_{N})_{\alpha,\eta}=\delta_{j,\eta}/\sqrt{m}
\>.
\end{equation} 
We restrict ourselves to disorder 
which under the similarity transformation yields
in the rotated basis a uniform wire as in 
Fig.~\ref{Fig:FindDragon}, 
thereby giving a quantum dragon nanodevice with ${\cal T}(E)=1$ 
\cite{NOVO2014}
and order amidst disorder for 
both the LDOS and the bond currents \cite{Novotny_2021,Novotny_2023}.  
Thus for constant $m$ in each slice the unitary matrix allows 
one to write the sufficient condition for a quantum dragon nanodevice 
to be 
\begin{equation}
\label{Eq:FindDragon:A:B}
\begin{array}{llcll}
{\bf A}_j & {\vec v}_{\rm dragon} & = & & {\vec 0} 
\\
& & {\rm and} & \\
{\bf B}_{j,j+1} & {\vec v}_{\rm dragon} & = & -1 & {\vec v}_{\rm dragon} \\
\end{array}
\end{equation}
for all slices $j$.  Hence even with disorder all blocks of the 
Hamiltonian have ${\vec v}_{\rm dragon}$ as a common eigenvector.  
The associated eigenvalue of zero for all ${\bf A}_j$ is because, as in 
\ref{CSaF_AppA}, we have set the thin lead wires to have on site energy zero.  
Similarly, we have set the hopping in the thin lead wires to $-1$, 
the common eigenvalue for all ${\bf B}_{j,j+1}$ blocks.  

\subsubsection{Finding dragons: Quantum dragon of Fig.~2.}
The utilization of the unitary transformation is 
sketched in Fig.~\ref{Fig:FindDragon} for a
strongly disordered 
zigzag 2D hexagonal nanodevice with $\ell$$=$$73$ and $m$$=$$16$.
The top of Fig.~\ref{Fig:FindDragon} shows the physical space nanodevice 
with the tight-binding parameters color and size coded.  
The sphere sizes are proportional to the 
on site energy values $\epsilon_{i,j}$, with the color code: 
(light gray, 0.7), (dark gray, 0.8), (red, 0.9), (cyan, 1.0), 
(white, 1.1), (maroon, 1.2), and (blue, 1.3).  
The exceptions are 
the yellow spheres that denote atoms with on site energy zero.  
\added[id=MAN]{All cylinders have radii proportional to $t_{i,j;i',j'}$, and 
are cyan.  
The bonds which connect the device to the leads 
have hopping strength $1$$/$$\sqrt{m}$.
All bonds in the leads, and all inter-slice bonds, have 
strength $t_{i,j;i',j'}$$=$$1$.
}   
Using the similarity transformation described above, 
the device Hamiltonian 
(a $\ell m\times \ell m =1168 \times 1168$ matrix) 
of the top physical 
space depiction is changed to a different basis depicted pictorially on the bottom. 
For select atypicial disorder, the Hamiltonian in the new basis is block diagonal, with 
the two blocks $\ell\times\ell=73\times73$ and 
$\ell(m-1)\times\ell(m-1)= 1095\times 1095$.  
The $\ell\times\ell$ block is a uniform 1D wire with $\ell=73$ atoms, 
and is the only block of the Hamiltonian which is connected to the leads.  
The $1095\times 1095$ block of the Hamiltonian is strongly disordered, 
but is disconnected from the 
leads and hence does not influence the electron transmission.  
The semi-infinite wires in physical space connect using hopping strengths 
${\vec v}_{\rm Dragon}$ to the atoms in the 
end slices of the nanodevice, and hence are 
only connected to the uniform wire segment in the new basis.  
The real-space device attached to leads (top) as well as the connected 
1D wire (bottom) 
both have ${\cal T}(E)$$=$$1$ for all $-2$$<$$E$$<$$2$, 
although intuitively only the bottom representation shows 
an obvious \lq short circuit' behavior.  

\subsubsection{Finding dragons: Specific example}

Consider a special case in order to illustrate the similarity transformation.  
This special case occurs for nanodevices based on 2D hexagonal 
(as in Fig.~\ref{Fig:FindDragon}) 
and 2D square-octagonal graphs of 
Fig.~\ref{Fig:2D:SqOct}.  
We consider graphs with $m$ vertices in every slice.  

Since every slice has $m$ atoms, all of the inter-slice matrices 
can be chosen to be 
${\bf B}_{j,j+1}=-{\bf I}_{m}$ with the  
$m\times m$ identity matrix ${\bf I}_{m}$.
Hence for all ${\bf B}_{j,j+1}$ matrices ${\vec v}_{\rm Dragon}$ is 
a common eigenvector with eigenvalue $-1$.  
This means the only inter-slice bonds have hopping strength 
\added[id=MAN]{one, namely} 
\begin{equation}
\label{Eq:tInterI}
t_{i,j;i',j'} \> = \> \delta_{i,i'} \> \delta_{j',j\pm 1} 
\>.
\end{equation}
The eigenvalue of ${\bf B}_{j,j+1}$ is the value of 
the hopping term in the attached leads, so the 
uniform wire in the rotated basis in Fig.~\ref{Fig:FindDragon} 
is the same as the attached leads 
(as in \ref{CSaF_AppA}, we have taken the 
hopping strength in the semi-infinite wires as our unit of energy). 
Note in Eq.~(\ref{Eq:HamTB}) there is a negative sign 
put in, the same as for a uniform 1D wire in \ref{CSaF_AppA}.  

Here we consider the case where every vertex has at most one 
intra-slice bond associated with it.  
For a 2D hexagonal device with only nn bonds (as in 
\added[id=MAN]{Figs.~\ref{Fig:FindDragon}, \ref{Fig:DragonAlmost}, 
\ref{Fig:DragonManyFano:zC}, and \ref{Fig:2D:SqOct})}, 
every atom is connected to at most one 
intra-slice bond.  
Whenever there is an intra-slice bond between atoms $i,j$ and $i+1,j$, 
we need to ensure the uniform ${\vec v}_{\rm Dragon}$ is 
an eigenvector of ${\bf A}_{j}$ with eigenvalue zero.   
Because there is at most one intra-slice bond per atom, 
we require for every intra-slice bond the condition 
\begin{equation}
\label{Eq:1BondDragon2}
\begin{array}{rcl}
\left(\begin{array}{cc}
\epsilon_{i,j} & -t_{i,j;i+1,j} \\
-t_{i,j;i+1,j} & \epsilon_{i+1,j} \\
\end{array}\right) 
\left(\begin{array}{c}
1 \\ 1 \\
\end{array}\right) 
& \> = \> & 
\left(\begin{array}{c}
0 \\ 0 \\
\end{array}\right) 
\\
{\rm which \> yields} & & \\
\epsilon_{i,j} \> = \> \epsilon_{i+1,j} & = & t_{i,j;i+1,j}
\>.
\end{array}
\end{equation}
We need the eigenvalue of ${\bf A}_j$ associated with 
${\vec v}_{\rm Dragon}$ to be zero in order to make 
the uniform wire in the rotated basis in Fig.~\ref{Fig:FindDragon} 
be the same as the attached leads of \ref{CSaF_AppA}, 
where we chose our zero of energy as 
the value of the on~site energy of the lead atoms.  
Equation~(\ref{Eq:1BondDragon2}) is satisfied for atypical disorder with 
this {\it local\/} correlation 
whenever an intra-slice bond is present. 

It is important to stress Eq.~(\ref{Eq:1BondDragon2}) has only 
{\it local\/} correlations. Each intra-slice bond strength can be 
chosen independently from every other one, and furthermore is 
independent of $m$ and of $\ell$.  
In other words, one can independently assign any arbitrarily 
value for $t_{i,j;i+1,j}$, make the two associated on~site energies 
satisfy the \lq dragon condition' of 
Eq.~(\ref{Eq:1BondDragon2}), and have a device with 
order amidst disorder and ${\cal T}(E)=1$.  
The intra-slice bond strengths 
$t_{i,j;i+1,j}$ can be chosen to be random, or can be 
chosen as in Fig.~\ref{Fig:FindDragon} to print a 
design.  

\section{Fano Resonances}
\label{sec:3:Fano}

\begin{center}
\begin{figure}[tb]
\includegraphics[width=0.33\textwidth]{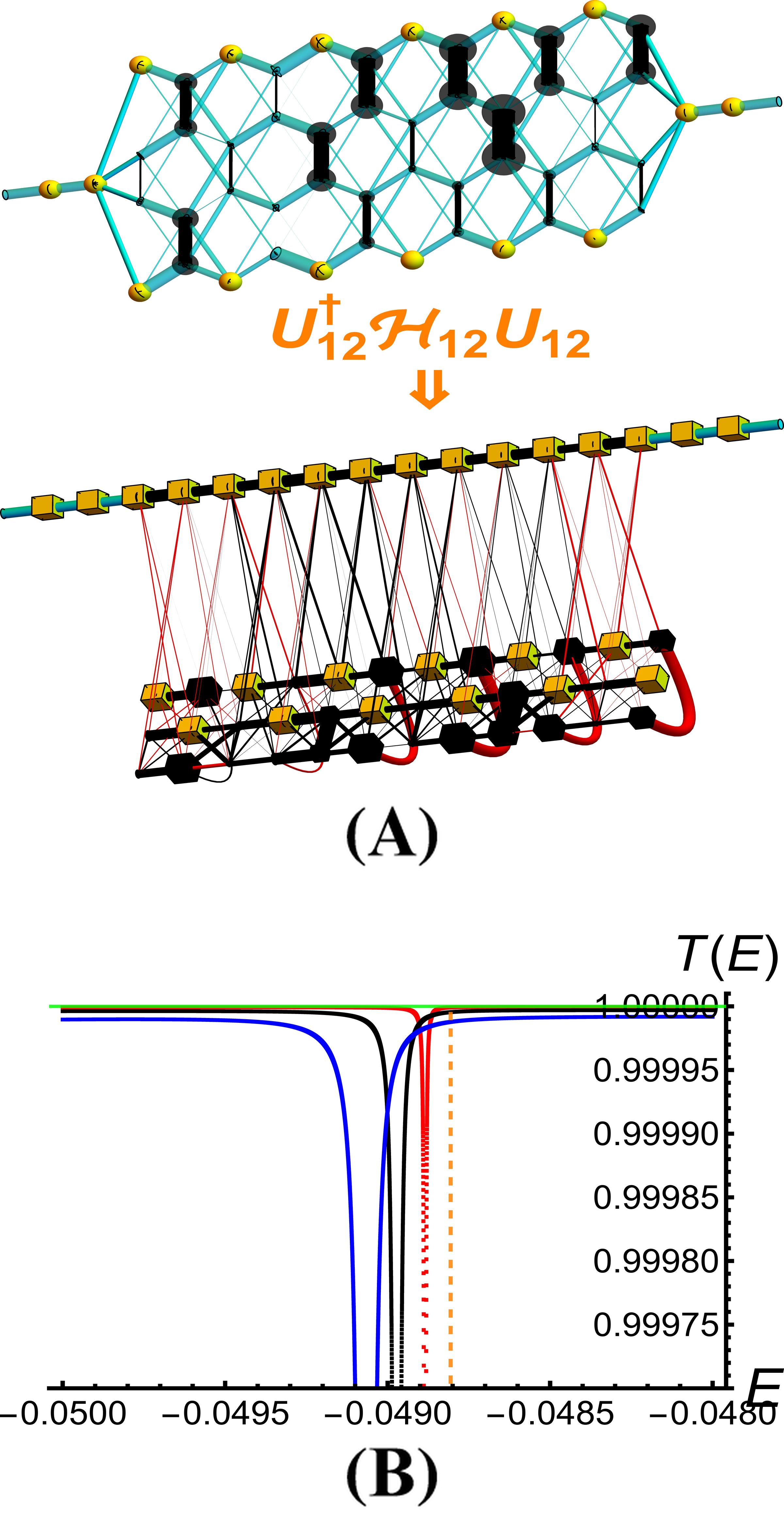}
\caption{
\label{Fig:DragonAlmost}
The effect of the Hamiltonian of Eq.~(\ref{Eq:HamTB}) almost being a quantum dragon.  
{\bf (A)}~ A device based on a 2D hexagonal graph with $\ell=12$ and $m=4$, 
showing the effect of the unitary transformation ${\bf U}_{12}$ on the device 
depiction when the device Hamiltonian is close to being a quantum dragon.  
{\bf (B)}~ Shows ${\cal T}(E)$ for 
an example of a similar quantum dragon nanodevice based on a 2D hexagonal graph 
with $\ell=80$ and $m=22$ for four different 
strengths of uncorrelated disorder, $\delta=0$ (green, ${\cal T}(E)=1$ for all $E$), 
$\delta=0.002$ (red), $\delta=0.004$ (black), and $\delta=0.006$ (blue).  
The dashed vertical orange line is the location of an eigenvalue of 
the $\delta=0$ device Hamiltonian.  
In agreement with Eq.~(\ref{Eq:Fano01}), 
all three finite $\delta$ values have ${\cal T}(E)$ that plunge to values 
numerically indistinguishable from zero.  Note the expanded scales 
for $E$ and ${\cal T}(E)$.  
See text in Sec.~\ref{sec:3:Fano} for full description.  
}
\end{figure}
\end{center}

It is natural to ask what the transmission ${\cal T}(E)$ will be for nanodevices 
which have the tight binding parameters in Eq.~(\ref{Eq:HamTB}) just slightly 
different from the tight binding parameters which give a 
quantum dragon nanodevice, i.e., 
what is the effect on a quantum dragon if the 
similarity transformation in 
Fig.~\ref{Fig:FindDragon} only approximately block diagonalizes the Hamiltonian.  
The same analysis will hold for nearly ballistic nanodevices which have zero disorder.  
This question is the main emphasis of this paper.  
The underlying effects are shown in Fig.~\ref{Fig:DragonAlmost}.  

Fig.~\ref{Fig:DragonAlmost}{\bf (A)}~shows a 2D hexagonal graph 
with $\ell=12$ and $m=4$, 
including both nn and nnn inter-slice bonds.  
The intra-slice bonds and 
on site energies satisfy Eq.~(\ref{Eq:1BondDragon2}), 
depicted as black vertical bonds and spheres.  
All nnn inter-slice bonds were chosen randomly in 
$t_{i,j;i\pm1,j+1}\in[0.0,0.2]$ 
(or zero if $i$$-$$1$$<$$1$ or $i$$+$$1$$>$$m$), 
and a quantum dragon nanodevice requires every nn 
inter-slice hopping strength between 
adjacent slices $j$ and $j$$+$$1$ to satisfy 
\begin{equation}
\label{Eq:InterDragon_NN_NNN} 
t_{i,j;i,j+1} \> = \> 1 \> - \> 
t_{i,j;i+1,j+1} \> - \> 
t_{i,j;i-1,j+1}
\end{equation}
in accordance with Eq.~(\ref{Eq:FindDragon:A:B}).
The unitary 
transformation ${\bf U}_{12}$ would then yield a uniform wire segment 
connected to the leads, with the uniform wire segment in the new basis 
having all on~site energies zero (the yellow cubes in the new basis) and 
hopping strengths uniform and equal to unity.  
Once the quantum dragon conditions have been met for the nanodevice, 
uncorrelated disorder may be added to give a nanodevice that is 
nearly a quantum dragon.  
In Fig.~\ref{Fig:DragonAlmost}{\bf (A)} the quantum dragon values for 
each inter-slice nn bonds $t_{i,j;i,j+1}$ from 
Eq.~(\ref{Eq:InterDragon_NN_NNN}) were 
randomly multiplied by a random number in $[1-\delta,1+\delta]$.  
Such added uncorrelated disorder gives a nanodevice which is almost a 
quantum dragon, and leads to weak hopping (with the sign of the hopping being either 
negative [black] or positive [red]) in the new basis between 
the sites which form the wire and the sites which are disconnected 
in a quantum dragon.  This disorder also introduces different strengths 
of hopping in the rotated basis between adjacent wire nodes. 
These weak hopping terms in the new basis between the wire sites and 
previously disconnected sites lead to Fano resonances in 
the transmission \cite{MIRO2010}.  

More accurately, as seen in Fig.~\ref{Fig:DragonAlmost}{\bf (B)}, 
there are Fano anti-resonances which suppress the transmission probability all 
the way to zero at particular energies.  
A Fano resonance \cite{FANO1961} in electron transmission \cite{MIRO2010} depends on an 
asymmetry parameter $q_{\rm Fano}$, a resonant energy $E_{\rm Fano}$, 
and a resonant width $\Gamma_{\rm Fano}$.  
The transmission can be given by \cite{MIRO2010}
\begin{equation}
\label{Eq:Fano01}
\begin{array}{c}
\begin{array}{lcl}
{\cal T}(E) & \> = \> & \frac{1}{1+q_{\rm Fano}^2} \> 
\frac{\left(\epsilon_{\rm Fano}+q_{\rm Fano}\right)^2}{\epsilon_{\rm Fano}^2+1}
\\ 
\\
\epsilon_{\rm Fano} & = & \frac{2\left(E-E_{\rm Fano}\right)^2}{\Gamma_{\rm Fano}}
\\
\end{array} 
\\
\\
{\rm with} 
\\
\\
{\rm extrema:} \>
\left\{\begin{array}{ccc}
{\cal T}_{\rm min}=0 & \> {\rm at} \> & \epsilon_{\rm Fano}=-q_{\rm Fano} \\
{\cal T}_{\rm max}=1 & \> {\rm at} \> & \epsilon_{\rm Fano}=1/q_{\rm Fano} \\
\end{array}\right.
\end{array}
\end{equation}
which is bounded between zero and 
unity\footnote{Unfortunately the literature typically uses 
$\epsilon_{i,j}$ for the 
tight binding on~site energy 
and $\epsilon_{\rm Fano}$ for the dimensionless distance 
from $E_{\rm Fano}$, but 
they should not be confused since they are differentiated by the subscripts.  
Similarly, $q_{\rm Fano}$ is not a wavevector.}. 
The locations of the (possible) resonance energies $E_{\rm Fano}$
for a material very close to being a quantum dragon 
with Hamiltonian ${\cal H}_{\rm Dragon}$ are given by 
the eigenvalues of ${\cal H}_{\rm Dragon}$ (at least within 
a tight binding approximation where the Hilbert space is 
of finite dimension).  For a device with a single Fano resonance, 
when $\left|E-E_{\rm Fano}\right|$ is large the transmission approaches 
${\cal T}\rightarrow 1-q_{\rm Fano}^2$ for small $q_{\rm Fano}$.  
When a very small amount of uncorrelated disorder is added to a quantum dragon, 
away from any given Fano resonance the transmission must approach unity
as seen in Fig.~\ref{Fig:DragonAlmost}{\bf (B)}.   
Hence a small but non-zero value for added disorder to a quantum 
dragon requires a small value for $q_{\rm Fano}$.  
Hence in Fig.~\ref{Fig:DragonAlmost}{\bf (B)} $q_{\rm Fano}\ll 1$, 
and the transmission goes to zero in line with 
Eq.~(\ref{Eq:Fano01}).  For the same quantum dragon nanodevice 
different variants from the distribution for the uncorrelated 
randomness will have different values of 
$E_{\rm Fano}$, $\Gamma_{\rm Fano}$, and $q_{\rm Fano}$.  
Hence for different variants ${\cal T}(E)$ goes to zero at 
different energies.  
Hence to calculate ${\cal T}_{\rm ave}(E)$ for a fixed energy 
using $M$ variants 
will require $M$ to be large.  

Physically the uncorrelated disorder may be from 
a combination of slight differences in the hopping strengths 
[as in Fig.~\ref{Fig:DragonAlmost}(A)], in the 
onsite energies (as considered in the subsequent sections), 
or due to finite temperature effects (quenched phonons).   
The uncorrelated disorder of any origin causes Fano resonances with widths 
related to its strength $\delta$
\cite{MIRO2010}.  
More accurately there are Fano anti-resonances which suppress the transmission probability all 
the way to zero at particular energies, while for very small $\delta$ 
away from these resonance energies the transmission is ${\cal T}(E)\approx1$ for all $E$.

\section{Transmission Averaged over Disorder: Approaching a Quantum Dragon}
\label{sec:4:NearQuantumDragon}

\begin{center}
\begin{figure}[tb]
\includegraphics[width=0.47\textwidth]{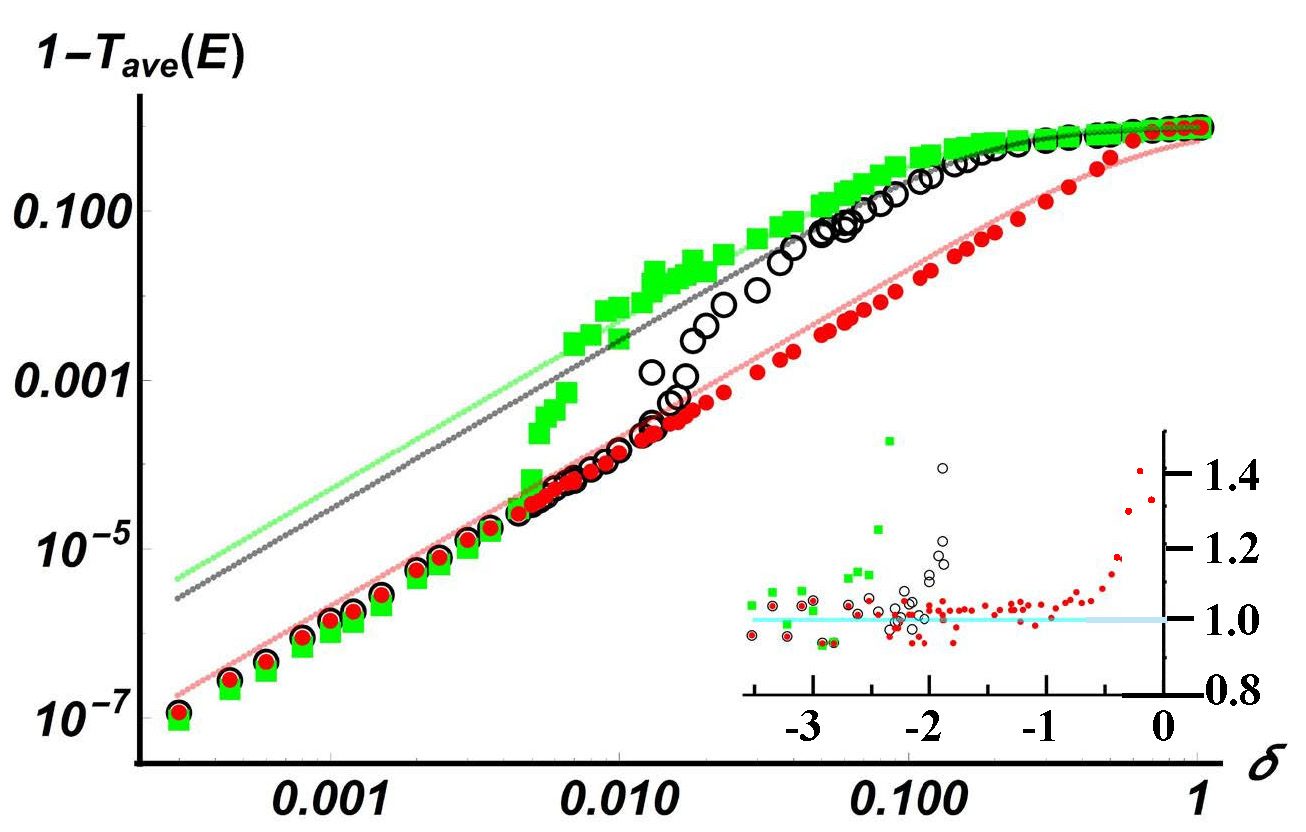}
\caption{
\label{Fig:DragonFewFano}
The quantity $1-{\cal T}_{\rm ave}(E)$ vs $\delta$ 
is shown for a single quantum dragon 
based on a 2D hexagonal graph 
with $\ell=80$ and $m=20$ so $N=1600$.  
The three energies shown are $E=-1,\>0,\>1$ shown as 
black, green, and red (open circles, squares, disks), respectively.  
The averages are over $M$$=$$10^5$ different values of the uncorrelated 
on site disorder 
chosen from a Gaussian distribution 
with mean zero and width $\delta$, with the random deviate 
chosen independently for each vertex and added to the 
on site energies of the quantum dragon values of that vertex.  
The three curves, color coded to the data points, 
show the predictions of Eq.~(\ref{Eq:T-scaling}) 
with no adjustable parameters.  
Asymptotically for small $\delta$ these solid curves are close to a 
straight line of slope~2, agreeing with 
Eq.~(\ref{Eq:scale:Sec4}).  
We see $1-{\cal T}_{\rm ave}$ is proportional to $\delta^2$, 
but exhibits a 
cross-over from averages dominated by a single nearby 
Fano resonance for small $\delta$ 
to the Eq.~(\ref{Eq:T-scaling}) prediction when the averages 
are due to many Fano resonances. 
{\bf Inset:}~ The same data illustrating the 
universal scaling predicted for very small $\delta$ in 
Eq.~(\ref{Eq:deltaSmall:01}), with the abscissa 
${\rm log}_{10}(\delta)$.   
The ordinate of each point is the lhs of 
Eq.~(\ref{Eq:deltaSmall:01}), while the rhs of 
Eq.~(\ref{Eq:deltaSmall:01}) is the 
horizontal cyan line.  
See text in Sec.~\ref{sec:4:NearQuantumDragon} for full description.  
}
\end{figure}
\end{center}

When averaging over $M$ 
variants of the uncorrelated disorder the sharp Fano anti-resonances with perfect 
suppression of the transmission ${\cal T}(E)$ 
as in Sec.~\ref{sec:3:Fano} 
get fuzzy and the transmission averaged over uncorrelated disorder 
${\cal T}_{\rm ave}(E)$ 
stays for a weak disorder strength $\delta$ close to $1$.  
See Fig.~\ref{Fig:DragonFewFano} for purely on site uncorrelated disorder 
at three energies for the same quantum dragon nanodevice. 
In Fig.~\ref{Fig:DragonFewFano} one can see that the three curves corresponding to 
different energies nearly collapse onto the same line for sufficiently weak disorder 
and only with increasing 
disorder strength $\delta$ they start to differ significantly.  
For the three energies shown in Fig.~\ref{Fig:DragonFewFano} there is then an 
energy dependent crossover to another scaling regime that holds for 
moderate values of $\delta$.  It is seen in both of these 
scaling regimes that 
\begin{equation}
\label{Eq:scale:Sec4}
1-{\cal T}_{\rm ave}(E) \> \propto \delta^2
\end{equation}
as shown by the 
log-log plot in Fig.~\ref{Fig:DragonFewFano}.  

We analyze analytically via perturbation theory both of these 
regimes where Eq.~(\ref{Eq:scale:Sec4}) holds.  We will find 
{\it universal scaling behavior\/} in both of these regimes.  

For the weakest disorder strength $\delta$ the universal scaling is derived in 
\ref{CSaF_AppB}.  
This universal behavior for the weakest disorder can be understood analytically by a 
lowest-order perturbation theory in the disorder strength $\delta$ 
and results in the universal Eq.~\eqref{Eq:AppB:final} where the only appearing dragon 
parameters are the nanodevice sizes $\ell$ and $m$. 
No other details of a particular dragon nanodevice, or ballistic nanodevice, enter. 
The energy dependence of this result is very weak away from the band edges, 
and is determined solely by the dispersion relation of uniform leads.   
This universal scaling of Eq.~(\ref{Eq:deltaSmall:01}) for weakest disorder is shown in the 
inset in Fig.~\ref{Fig:DragonFewFano}, which plots ${\rm log}_{10}(\delta)$ 
versus the scaling prediction of Eq.~(\ref{Eq:deltaSmall:01}) 
which 
should be equal to unity (the cyan horizontal line).  
Since $E=1$ is in a range with a very small DOS$(E)$ the 
Eq.~(\ref{Eq:deltaSmall:01}) 
universal scaling holds up to about $\delta\approx 0.2$.  
For the other energies shown in Fig.~\ref{Fig:DragonFewFano} the cross over 
from this very small $\delta$ universal scaling happens for smaller 
$\delta$ due to the larger DOS$(E)$ for those energies.  
We numerically test the predicted universal scaling in this small $\delta$ regime in 
Sec.~\ref{sec:5:SmallDelta} for different 2D ballistic 
nanodevices and quantum dragon nanodevices.  

With increasing disorder strength $\delta$ the situation dramatically 
changes and the transmission becomes strongly energy dependent. 
This corresponds to the regime of many Fano anti-resonances 
being important near an energy $E$.  
This scaling regime 
can be described by a generalization of the 
scaling approach of Ref.~\cite{MAR2007} used for 
an analysis of doped silicon nanowires 
that have nearly ballistic electron transmission.  
We have generalized this analysis to include quantum dragon nanodevices 
in \ref{CSaF_AppC} and 
derive the second regime of universal scaling where 
Eq.~(\ref{Eq:scale:Sec4}) holds.  
We numerically test the predicted universal scaling in this second regime in 
Sec.~\ref{sec:6:ScaleDOS} for different 2D quantum dragon nanodevices.  

The summary, the two different universal scaling regimes have 
\begin{equation}
\label{Eq:scale:Sec4:summary}
1 \! - \! {\cal T}_{\rm ave}(E) = \!
\left\{
\!
\begin{array}{lclcl}
\frac{\delta^2}{4-E^2} \frac{\ell}{m} 
& \>
& {\rm very\> small\>} \delta 
\\
\>
\\
\frac{\Upsilon \delta^2}{1+\Upsilon\delta^2}
\approx 
\Upsilon \> \delta^2 
& & {\rm intermediate\>} \delta
\end{array}
\right.
\end{equation}
with the definition 
$\Upsilon \> = \> 2 \pi L_{\rm scale} \> {{\rm DOS}(E)} \> \! \Big/ {\sqrt{4-E^2}}$.    
The last expression for intermediate $\delta$ is only valid when 
$\delta^2 \Upsilon \ll 1$, 
and is shown by the approximate slope$=$$2$ in the curves in 
Fig.~\ref{Fig:DragonFewFano}.  
For even larger $\delta$ the nanodevice would be in the diffusive 
regime, with ${\cal T}_{\rm ave}(E)$ small
and $L_e\ll L \ll \xi_{\rm A}$,  wherein the 
predictions of the Dorokhov-Mello-Pereyra-Kumar theory 
\cite{Dorokhov1982,Mello1988} for the electrical conductance 
would hold.

\section{Universal Scaling for small $\delta$}
\label{sec:5:SmallDelta}
\ref{CSaF_AppB} derives using a Dyson series method the expected 
average transmission ${\cal T}_{\rm ave}(E)$, averaged 
over uncorrelated random on site disorder of strength 
$\delta$.  
The random uncorrelated disorder is added to the 
on site energies of the quantum dragon or translationally 
invariant ballistic device.  
Rewriting from the final expression, 
Eq.~(\ref{Eq:AppB:final}), gives 
the predicted scaling relation 
\begin{equation}
\label{Eq:deltaSmall:01}
\begin{array}{lcl}
\left[1-{\cal T}_{\rm ave} (E)\right]
\>\frac{(4-E^{2})}{\delta^{2}}
\> \frac{m}{\ell}
& \>=\> & 1
\>. \\
\end{array}
\end{equation}
The inset of Fig.~\ref{Fig:DragonFewFano}
gives one illustration of the scaling of Eq.~(\ref{Eq:deltaSmall:01}) 
for a single quantum dragon nanodevice.  

\begin{center}
\begin{figure}[tb]
\includegraphics[width=0.47\textwidth]{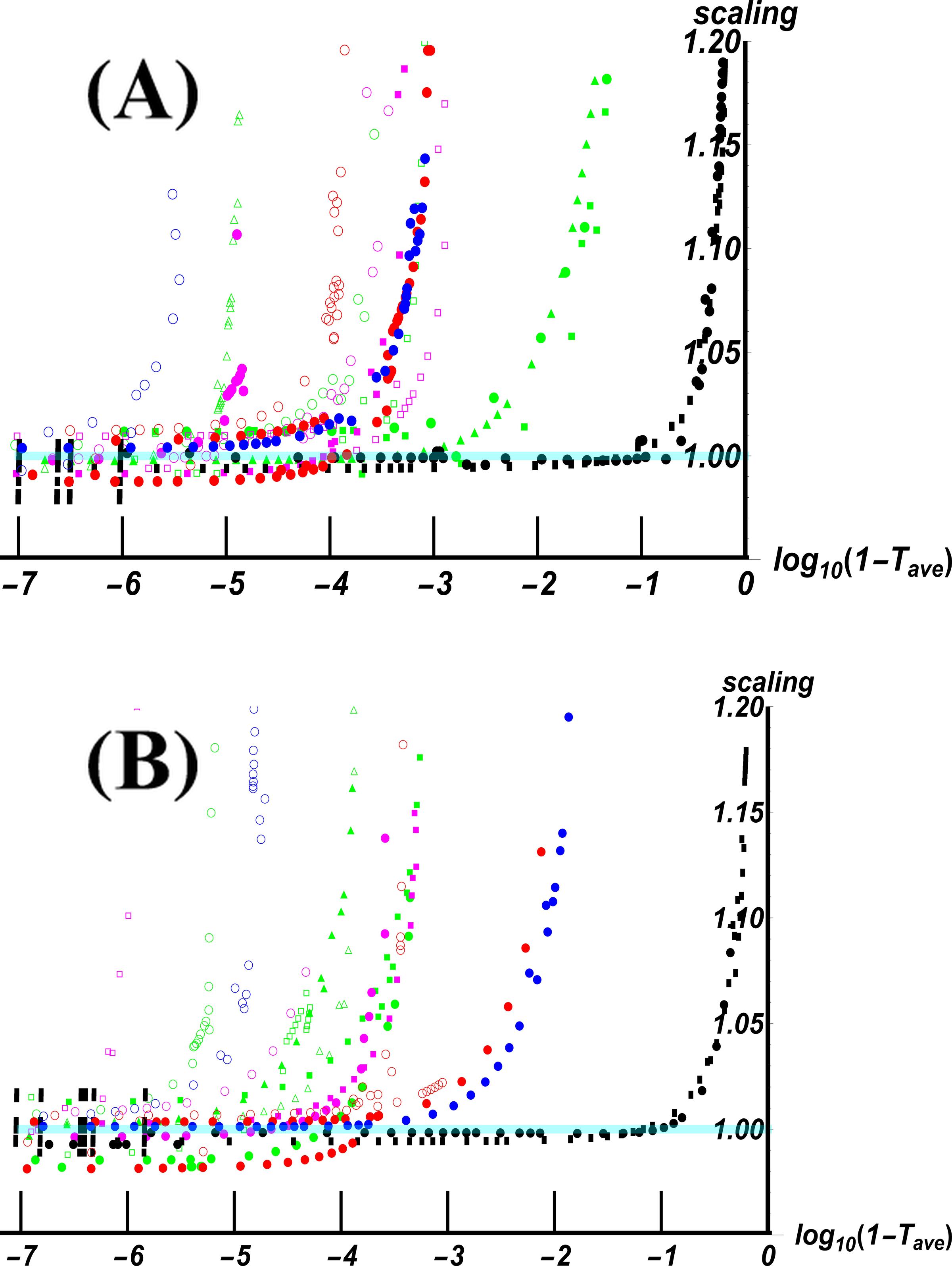}
\caption{
\label{Fig:Small:delta} 
A test of the small~$\delta$ universal scaling of 
Eq.~(\ref{Eq:deltaSmall:01}),
for five different types of nanodevices 
depicted by the five different colors.  
The predicted scaling value of unity is shown by the horizontal cyan lines.  
The energies shown are {\bf (A)} $E=1$ and 
{\bf (B)} $E=-\sqrt{2}$.  
See Sec.~\ref{sec:5:SmallDelta} and \ref{CSaF_AppB} 
for complete information.  
}
\end{figure}
\end{center}

See Fig.~\ref{Fig:Small:delta} for a test of 
the universal scaling of 
Eq.~(\ref{Eq:deltaSmall:01}) starting from multiple 
ballistic nanodevices and quantum dragon nanodevices.  
Note there are no adjustable parameters in the 
scaling plot.  
The data are for five types of nanodevices:
\begin{itemize}
\item[$\bullet$] {\bf Green symbols:}~ 
Single-walled armchair nanotubes with ballistic electron propagation 
for three values of ($m,\ell$).  
\begin{itemize}
\item[$\circ$] With $(m,\ell)$ given by $(12,37)$:disks, $(6,37)$:squares, and 
$(12,18)$:triangles.  
\end{itemize}
\item[$\bullet$] {\bf Magenta symbols:}~ 2D+3D quantum dragon 
nanodevice with constant $m$, as in 
Fig.~\ref{Fig:2D3DmConstant} with two values of $(\ell,m)$.  
\begin{itemize}
\item[$\circ$] With $(m,\ell)$ given by $(18,32)$:disks and $(10,32)$:squares.  
\end{itemize}
\item[$\bullet$] {\bf Blue symbols:}~ 3D quantum dragon nanodevice 
with constant $m$, as in 
Figure~\ref{Fig:3D:snake}.  
\begin{itemize}
\item[$\circ$] With $(m,\ell)$ given by $(18,20)$:disks.  
\end{itemize}
\item[$\bullet$] {\bf Red symbols:}~ 2D+3D nanodevice as in 
\cite{Novotny_2021}
with varying values of $m$ in each slice and hence using $m_{\rm Scale}$ from 
Eq.~(\ref{Eq:mScale}) for $m$ in Eq.~(\ref{Eq:deltaSmall:01}).  
\begin{itemize}
\item[$\circ$] Here $m\ell=N=372$ given by the open disks.  
\end{itemize}
\item[$\bullet$] {\bf Black symbols:}~ Linear chains (ballistic transmission 
in uniform metallic wires) with $m=1$ and two values of $\ell$.  
\begin{itemize}
\item[$\circ$] With $(m,\ell)$ given by $(1,37)$:disks, $(1,18)$:squares.  
\item[$\circ$] For $\ell=18$ at the smallest ${\cal T}_{\rm ave}(E)$ values 
are shown 
ten different averages, each using $M$$=$$10^4$ configurations of 
uncorrelated site disorder.  
\end{itemize}
\end{itemize}
In all cases in Fig.~\ref{Fig:Small:delta} 
the filled plotting symbols are for 
regular graphs and pure (not random) $\epsilon_{i,j}$ and $t_{i,j;i',j'}$, 
while the open plotting symbols are for corresponding 
graphs with 20\% of the intra-slice bonds cut and 
the other intra-slice hopping terms $t_{i,j;i',j}\in[0,2]$.  
In all cases the dragon condition 
\added[id=MAN]{Eq.~(\ref{Eq:FindDragon:A:B}), 
the same as 
Eq.~(\ref{Eq:2D3D:fixed_m}),}
is used to obtain the $\epsilon_{i,j}$ 
on site energy values. The averages are over $M$$=$$10^4$ Gaussian-distributed 
uncorrelated values added to each dragon condition on site energy.  
The Gaussian distribution has mean zero and width unity, and is then 
multiplied by $\delta$.  The averages must be over a large number 
$M$ of 
disorder configurations, due to the narrow Fano resonance dips 
to ${\cal T}(E)=0$ 
seen in Fig.~\ref{Fig:DragonAlmost}{\bf (B)}.  
The same dragon Hamiltonians and graphs are plotted for all five 
types of nanodevices, for 
the two energies {\bf (A)} $E=1$ and {\bf (B)} $E=-\sqrt{2}$.  
With no adjustable parameters, in Fig.~\ref{Fig:Small:delta} we see 
for small $\delta$ excellent agreement with 
Eq.~(\ref{Eq:deltaSmall:01}) 
for all ballistic nanodevices and quantum dragon nanodevices.  

The only prediction we do not provide is when 
for a particular type of nanodevice the universal scaling breaks 
down as $\delta$ becomes larger.  This is probably a complicated 
function of $\ell$, $m$, DOS$(E)$, $E$, and perhaps the precise location of the 
disorder in the quantum dragon nanodevice.  
In Fig.~\ref{Fig:Small:delta}{\bf (A)} the energy $E=1$ shown is near a minimum 
for the DOS$(E)$ for the single-walled nanotubes (green points).  
For a given quantum dragon nanodevice the universal scaling breaks 
down as $\delta$ increases at different values of $\delta$, 
and hence at different values 
of ${\cal T}_{\rm ave}(E)$, at different energies as seen most 
readily by comparing the SWNT (green) and 3D (blue) symbols in 
{\bf (A)} and {\bf (B)}.  
We observed very similar behavior for all energies we tested in 
$-2<E<2$, with the only difference from Fig.~\ref{Fig:Small:delta}
how large $\delta$ had to be before the universal scaling 
of Eq.~(\ref{Eq:deltaSmall:01}) 
broke down.  
This dependence of the scaling breakdown depending on 
$E$ was also observed in the inset of Fig.~\ref{Fig:DragonFewFano} 
for a different quantum dragon nanodevice.

\section{Universal Scaling related to DOS}
\label{sec:6:ScaleDOS}
In this section we test the universal scaling predicted 
in \ref{CSaF_AppC} wherein the DOS$(E)$ enters the scaling.  
In the three subsections below, 
we restrict our tests to 2D quantum dragons formed from 
nanoribbons.  However, it is important to realize that 
the universal scaling prediction is valid for any 
system with ballistic transport and also for any 
quantum dragon with any embedding dimension.  

We rewrite the universal scaling 
result of Eq.~(\ref{Eq:AppC:T-scaling})
as 
\begin{equation}
\label{Eq:T-scaling}
\begin{array}{lcl} 
``\mathrm{DOS}(E)" & = & 
\frac{\sqrt{4-E^2}}{2\pi\delta^2 L_{\rm scale}}
\left(\frac{1}{{\cal T}_{\rm ave}(E)}-1\right)
\\
\>
\\
& = & \frac{\sqrt{4-E^2}}{2\pi\delta^2 L_{\rm scale}}
\>\> \frac{1-{\cal T}_{\rm ave}(E)}{{\cal T}_{\rm ave}(E)}. 
\end{array}
\end{equation}
Here the quotes on the ``DOS$(E)$" $\>$mean the scaling 
given by a right-hand side (rhs) of Eq.~(\ref{Eq:T-scaling}) 
is an estimate under the assumptions required to obtain 
this scaling equation.  
Hence once we measure the average transmission 
${\cal T}_{\rm ave}(E)$ we can compare the universal scaling 
result from the rhs of Eq.~(\ref{Eq:T-scaling}) with the 
calculated DOS$(E)$ from our box counting algorithm for the 
Hamiltonian.  Note the DOS$(E)$ that enters is for the 
unperturbed Hamiltonian of the device (without leads) for 
either the ordered ballistic nanodevice or quantum dragon 
nanodevice.

\subsection{Quantum Dragons based on a 2D Hexagonal Graph} 
\begin{center}
\begin{figure*}[tb]
\includegraphics[width=0.97\textwidth]{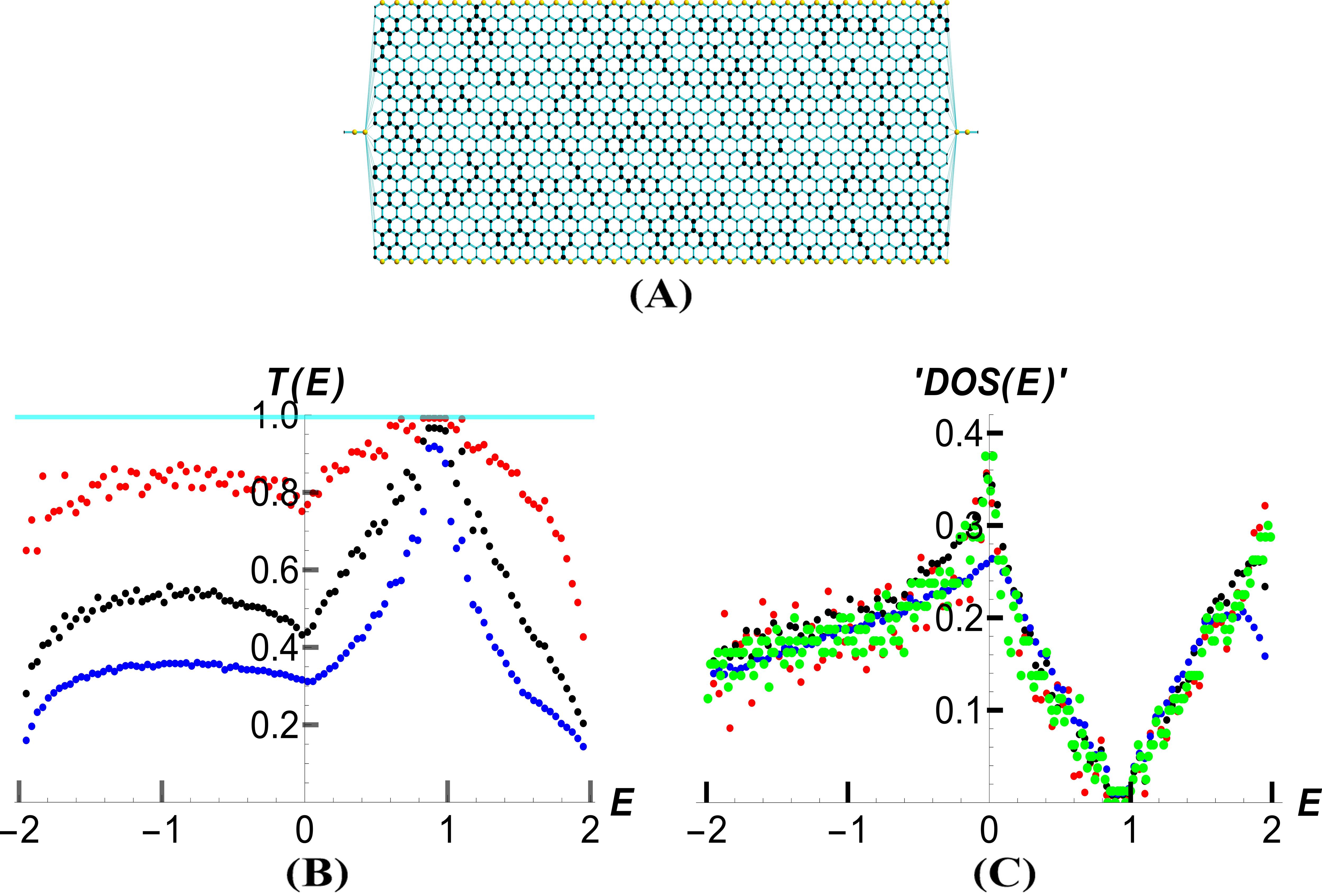}
\caption{
\label{Fig:DragonManyFano:zC}
\added[id=MAN]{A 2D hexagonal graph for an Eq.~(\ref{Eq:HamTB}) Hamiltonian 
with $m$$=$$20$ and $\ell$$=$$80$, 
which is a quantum dragon with ${\cal T}(E)$$=$$1$.  
The cyan cylinders have a radius proportional to the hopping 
strength $t_{i,j;i'j'}$.  The black spheres have a radius proportional to 
the on site energy $\epsilon_{i,j}$, 
while the yellow spheres have $\epsilon_{i,j}$$=$$0$.  
Only two atoms of each semi-infinite lead are shown.  
{\bf (B)}~ shows the average transmission ${\cal T}_{\rm ave}(E)$ versus 
energy for four values of additional uncorrelated disorder, 
$\Delta=0,\> 0.08,\>0.16,\> 0.24$ 
shown as cyan (horizontal line), red, black, and blue, respectively.} 
{\bf (C)}~ Shows the same data as {\bf (B)} scaled as in Eq.~(\ref{Eq:T-scaling}) 
for red, black, and blue symbols.  The green is the result of a box-counting 
of the average DOS$(E)$ from quantum dragon Hamiltonians ($\Delta$$=$$0$) 
with $10^4$ different $t_{i,j;i+1,j}=\left[0.5,1.5\right]$ realizations.  
The scaling uses $L=\sqrt{3}\ell a/2$ and 
$\delta^2=2\Delta^2/3$, so there are no adjustable parameters.  
See text in Sec.~\ref{sec:6:ScaleDOS}.1
for a more full description.  
}
\end{figure*}
\end{center}

\added[id=MAN]{Figure~\ref{Fig:DragonManyFano:zC}{\bf (A)}~ 
shows a quantum dragon nanodevice based on a 2D hexagonal graph 
with $m$$=$$20$ and $\ell$$=$$80$.  
Such nanodevices based on a 2D zigzag hexagonal ribbon 
as the underlying graph are similar to the one in Fig.~\ref{Fig:FindDragon}.    
Here the intra-slice bonds are randomly chosen with 
$t_{i,j;i+1,j}$$=$$\left[0.5,1.5\right]$ and associated on site energies 
chosen to satisfy $t_{i,j;i+1,j}$$=$$\epsilon_{i,j}$$=$$\epsilon_{i+1,j}$ 
as in Eq.~(\ref{Eq:1BondDragon2}), hence 
${\cal T}(E)$$=$$1$ for all $-2$$<$$E$$<2$.  
As there are no cut bonds, the underlying graph (not shown) is 
an ordered 2D hexagonal graph.  
}

Figure~\ref{Fig:DragonManyFano:zC}{\bf (B)}~ 
shows the average transmission ${\cal T}_{\rm ave}(E)$ for such quantum 
dragon nanodevices for three different non-zero values of $\delta$.  
\deleted[id=MAN]{The nanodevice is based on a 2D zigzag hexagonal ribbon 
with $\ell=80$ slices and $m=20$, 
so the underlying graph is similar to the one in Fig.~\ref{Fig:FindDragon}.    
The intra-slice bonds are randomly chosen with 
$t_{i,j;i+1,j}=\left[0.5,1.5\right]$ and associated on site energies 
chosen to satisfy $t_{i,j;i+1,j}=\epsilon_{i,j}=\epsilon_{i+1,j}$ 
as in Eq.~(\ref{Eq:1BondDragon2}), hence} 
Also shown is the $\delta$$=$$0$ result 
${\cal T}(E)=1$ for each nanodevice, 
presented as the cyan horizontal line.  
Each nanodevice has added 
uncorrelated disorder 
to each on~site energy of the quantum dragon Hamiltonian an 
additional random uniformly-chosen value in 
$[-\Delta_\epsilon,\Delta_\epsilon]$, 
and for the present intra-slice bonds an additional 
value chosen uniformly 
at random with $[-\Delta_t,\Delta_t]$.  

The analysis leading to Eq.~(\ref{Eq:T-scaling}) is only for 
random on site energies added to the dragon value, hence 
at first sight the analysis would have to be redone to take 
into account randomness in the bonds.  However, because there 
is at most one intra-slice bond per atom, no matter 
the random value of the intra-slice bond it is possible to have 
the on site energy set to the Eq.~(\ref{Eq:1BondDragon2}) dragon condition 
$\epsilon_{i,j}=t_{i,j;i+1,j}=\epsilon_{i+1,j}$.  
What enters Eq.~(\ref{Eq:T-scaling}) 
is only the variance $\delta^2$ of the uncorrelated disorder.  We need 
to relate the variance $\delta^2$ with the variance from the dragon 
condition.  The uniform probability distributions are 
${d\kappa_t}/{2\Delta_t}$ and 
${d\kappa_\epsilon}/{2\Delta_\epsilon}$.
This gives the calculation for atoms with an intra-slice bond 
of the variance from the dragon condition 
\begin{equation}
\label{Eq:Sec6:deltaVSkappa}
\frac{1}{4 \Delta_t\Delta_\epsilon}
\int_{-\Delta_t}^{\Delta_t} d\kappa_t 
\int_{-\Delta_t}^{\Delta_\epsilon} d\kappa_\epsilon 
\left(\kappa_t-\kappa_\epsilon\right)^2 
= \frac{\Delta_\epsilon^2+\Delta_t^2}{3}
\>.
\end{equation}
We choose $\Delta_t$$=$$\Delta_\epsilon$$=$$\Delta$, and so what enters 
into Eq.~(\ref{Eq:T-scaling}) is $\delta^2=2\Delta^2/3$.  
Furthermore, due to the hexagonal nature of the graph 
one has $L=\sqrt{3}\ell a/2$.  (These values are the lowest order terms as
they neglect edge effects, but should be appropriate 
for comparison with our statistics.)  
Figure~\ref{Fig:DragonManyFano:zC}{\bf (B)}~ shows the scaling of the same data as 
{\bf (A)} using Eq.~(\ref{Eq:T-scaling}). 
Here with no adjustable parameters since we use 
$L_{\rm scale}=L$, Figure~\ref{Fig:DragonManyFano:zC}{\bf (B)} 
shows excellent agreement with Eq.~(\ref{Eq:T-scaling}), 
with the three chosen values 
$\Delta=0.08,\>0.16,\> 0.24$ shown as red, black, blue, respectively. 
The transmission at each point is the average over 
$M$$=$$10^4$ different realizations 
of the uncorrelated 
on site parameters added to the 
quantum dragon value at that site. 
Also shown as green symbols in Fig.~\ref{Fig:DragonManyFano:zC}{\bf (B)}~ is a box 
counting result of the average ${\rm DOS}(E)$ per site from $10^4$ 
such $\Delta=0$ random quantum dragon Hamiltonians. 
This scaling should be reasonably good for any quantum dragon 
nanodevice in an intermediate regime of $\delta$.  
It should be reasonable for $\delta$ values smaller than when $L_{\cal T}(E)$ of 
Eq.~(\ref{Eq:L-T}) is 
comparable to or larger than $L/2$, which some of the blue symbols in 
Fig.~\ref{Fig:DragonManyFano:zC}{\bf (B)} may be just on the border of satisfying.  
It should also be reasonable if $\delta$ is not too small, in which case 
the universal scaling of Eq.~(\ref{Eq:deltaSmall:01}) is applicable.  
Note, for a pure armchair SWCNT (Single-Walled Carbon Nanotube) 
or zigzag CNR (Carbon Nanoribbon) 
$L=\sqrt{3}\left(\ell-1\right) a/2$ 
relates the physical device length $L$
to the number of slices $\ell$ in the underlying graph. 
For our quantum dragon nanodevices, we also use 
$L=\sqrt{3}\left(\ell-1\right) a/2$ and $L_{\rm scale}=L$ in 
the Eq.~(\ref{Eq:T-scaling}) scaling.  

A regular pure (zero disorder) armchair SWCNT or 
zigzag CNR has translational 
invariance (at least along the direction of current flow), 
and hence have been analyzed using band structure and Bloch wavefunction 
techniques.  
However, both can also 
be analyzed with the described similarity transformation in 
Sec.~\ref{sec:2:ModelMethod}, in particular for the special case 
in Sec.~\ref{sec:2:ModelMethod}.3.  
Nature has regular pure armchair SWCNTs and zigzag CNRs satisfying 
Eq.~(\ref{Eq:1BondDragon2}) for device on~site values $\epsilon_{i,j}=1$ 
since all bonds (both inter-slice and intra-slice) 
are identical and can be chosen to have 
strength $t_{i,j;i',j'}=1$ (by choosing the hopping strength of the attached 
semi-infinite leads to be the 
actual carbon-carbon bond hopping strength).  Thus the method of 
analysis sketched in Fig.~\ref{Fig:FindDragon} is an alternative 
analysis to obtain ballistic transport in 
regular pure armchair SWCNTs and zigzag CNRs.  
\added[id=MAN]{
Such analysis may also be related to some novel built nanoporous graphene \cite{Qin2024nanoporeGraphene} nanodevices. 
}

\subsection{Quantum Dragons based on a 2D Rectangular Graph}

\begin{center}
\begin{figure*}[tb]
\includegraphics[width=0.97\textwidth]{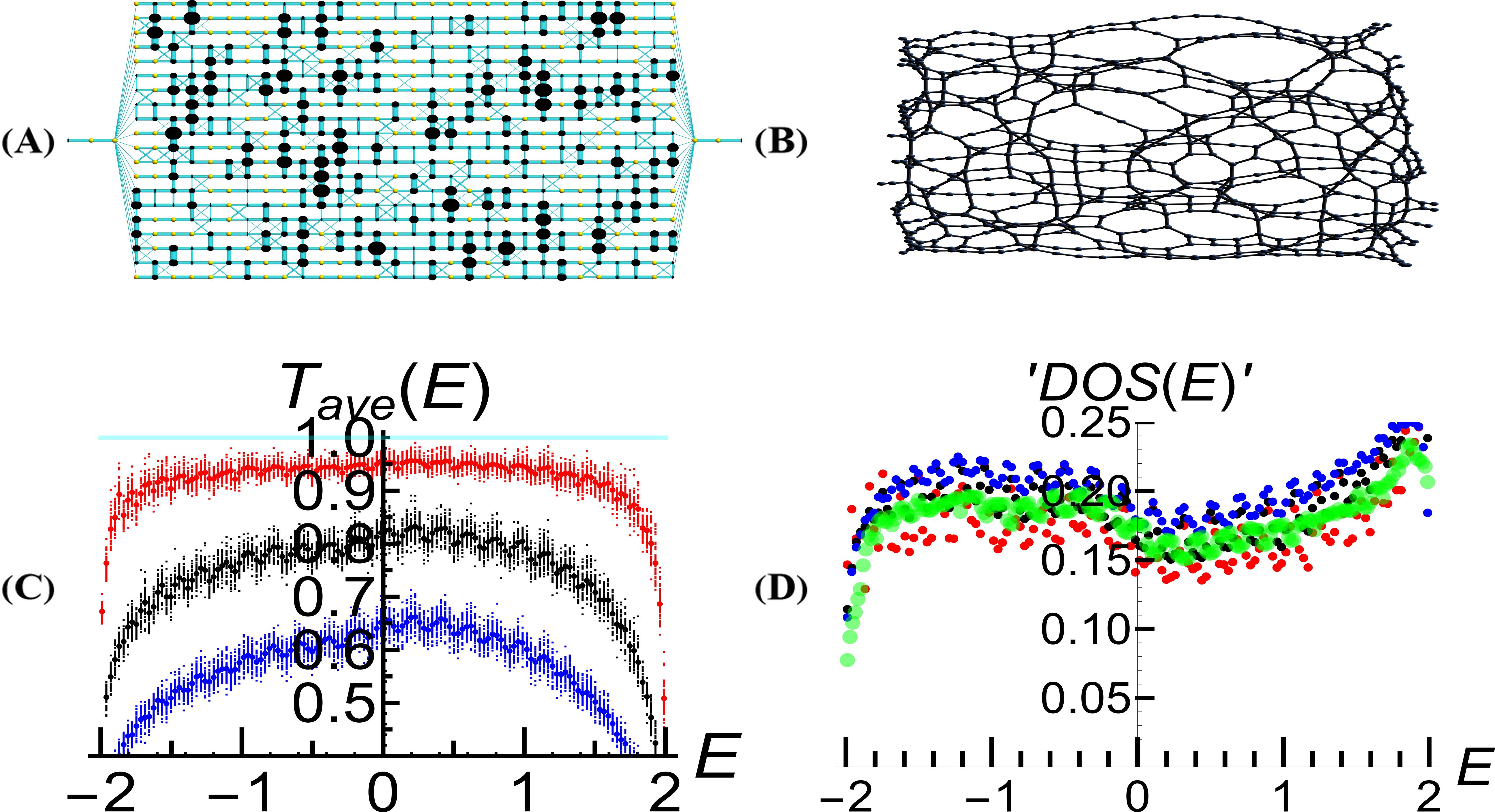}
\caption{
\label{Fig:2D:Rectangular:Scale}
Strongly disordered 
$\ell=30$ and $m=20$ nanoribbons based on 
2D rectangular graphs.  
(A) Shows an example of the disordered quantum dragon nanodevice, 
while (B) shows the underlying disordered graph of (A).
(C) Presents the transmission ${\cal T}(E)$ for the quantum 
dragon nanodevices (horizontal cyan line at ${\cal T}(E)=1$), as 
well as the average transmission ${\cal T}_{\rm ave}(E)$ for 50 different 
$\ell=30$ and $m=20$ quantum dragon nanodevices for three different 
values of the uncorrelated site disorder.  
(D) Exhibits the scaling of Eq.~(\ref{Eq:T-scaling}) 
${\cal T}_{\rm ave}(E)$ averaged over both the uncorrelated 
disorder and the 50 different quantum dragon nanodevices, 
using $L_{\rm scale}=0.55 L$.  Also shown is the DOS$(E)$ (green points) 
averaged over the 50 quantum dragons.  
See text Sec.~\ref{sec:6:ScaleDOS}.2 for a full description.  
}
\end{figure*}
\end{center}

Figure~\ref{Fig:2D:Rectangular:Scale} shows results for 50 quantum dragon nanodevices 
based on a 2D rectangular graph with $\ell=30$ and $m=20$.  
The intra-slice nn hopping strengths were chosen completely at random, such that 
50\% of the bonds were cut and the remaining bond strengths were chosen independently, 
uniformly at random with $t_{i,j;i+1,j}\in[0.2,1.8]$.  
The on~site energies were chosen 
for the device to be a quantum dragon from 
Eq.~(\ref{Eq:FindDragon:A:B}) 
with a uniform ${\vec v}_{\rm Dragon}$. 
The nnn inter-slice hopping strengths were chosen randomly such that 20\% of the nnn bonds were 
present with $t_{i,j;i\pm1,j+1}\in[0.1,0.4]$ and the nn inter-slice hopping 
strengths chosen 
from Eq.~(\ref{Eq:FindDragon:A:B}) to make the device a quantum dragon.  
The disorder hence has only local correlations, but each device shows 
order amidst disorder.  
Figure~\ref{Fig:2D:Rectangular:Scale}{\bf (A)} shows an 
example of one such quantum dragon nanodevice, 
while another view of the underlying graph is 
Fig.~\ref{Fig:2D:Rectangular:Scale}{\bf (B)} 
using the {\tt Mathematica\/} function {\tt GraphPlot} \cite{Mathematica}.  
Figure~\ref{Fig:2D:Rectangular:Scale}{\bf (C)}~ shows the average transmission over 
$M$$=$$10^3$ different 
realizations of strength $\delta$ of uncorrelated on~site disorder, chosen 
from a Gaussian distribution of mean zero and width unity uniformly and then 
multiplied by $\delta$.  The four values of $\delta$ are $0.0$ 
(cyan, quantum dragon value), 
$0.08$ (red), $0.16$ (black), and $0.24$ (blue).  Each energy point plotted was run for 50 different 
quantum dragon nanodevices constructed with the type of disorder described 
above and all 50 are plotted, 
together with their average (larger symbol size).  
The quantum dragon result, found for all 50 nanodevices, 
$\delta=0$ is shown in cyan with ${\cal T}(E)=1$.  
Figure~\ref{Fig:2D:Rectangular:Scale}{\bf (D)}~ shows the same finite $\delta$ 
values with the same 
color codes for the average transmissions in Figure~\ref{Fig:2D:Rectangular:Scale}{\bf (C)}, 
scaled according to Eq.~(\ref{Eq:T-scaling}), 
using $L_{\rm scale}=0.55L$.    
The dark green points show the result for the DOS for all 50 device Hamiltonians, 
with a box counting 
using $\Delta E=0.15$. The agreement between the box counting DOS$(E)$ 
and ``DOS$(E)$" ~from Eq.~(\ref{Eq:T-scaling})
is extremely good.  

\subsection{Quantum Dragons based on a 2-D Square-Octagonal Graph}

\begin{center}
\begin{figure*}[tb]
\includegraphics[width=0.97\textwidth]
{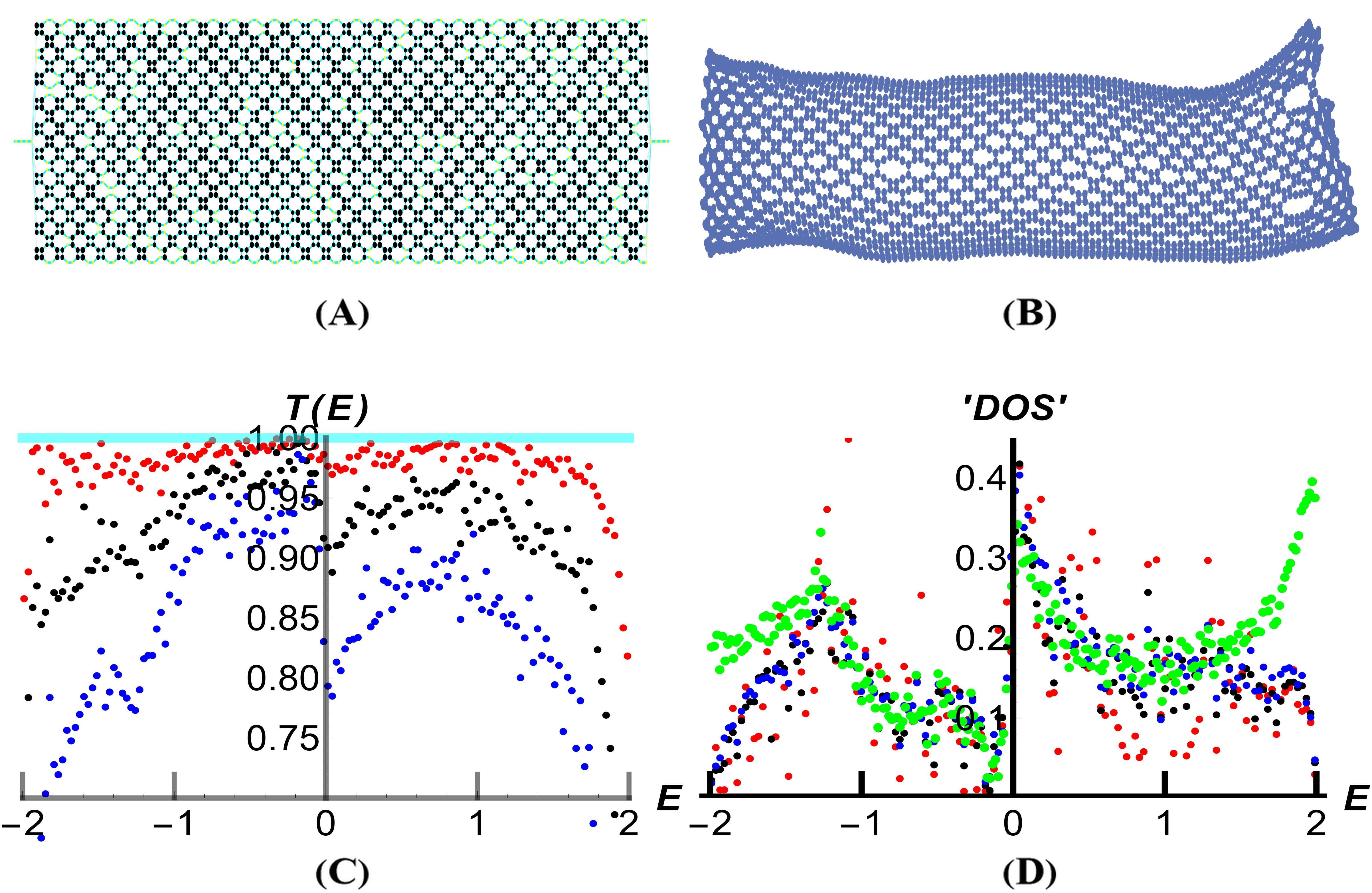}
\caption{
\label{Fig:2D:SqOct}
A strongly disordered 
$\ell=120$ and $m=20$ quantum dragon nanodevice based on a 
square-octagonal graph.  
(A) Shows the nanodevice, with the disordered graph exhibited in (B). 
(C) Presents the quantum dragon transmission ${\cal T}(E)$ as the 
light cyan 
horizontal line, as well as the average transmission 
${\cal T}_{\rm ave}(E)$ for three values of uncorrelated on site disorder
\added[id=MAN
]{ for $\delta=0.02$ (red), $\delta=0.04$ (black), $\delta=0.06$ (blue).} 
(D) Shows the scaling using Eq.~(\ref{Eq:T-scaling}) of the data in (C), 
together with the DOS$(E)$ (green points).  
Here the scaling has one adjustable parameter, using 
$L_{\rm scale}=70$ rather than the device length $L=101$. 
See text Sec.~\ref{sec:6:ScaleDOS}.3 for a full description. 
}
\end{figure*}
\end{center}

Fig.~\ref{Fig:2D:SqOct} shows results for a disordered 
square-octagonal graph.  Every intra-slice bond, and 
the associated on site energies, must satisfy 
Eq.~(\ref{Eq:1BondDragon2}) 
in order for the nanodevice to exhibit 
order amidst disorder and having complete 
electron transmission for all energies $-2$$\le$$E$$\le$$2$.  
If an intra-slice hopping is chosen to be 
$t_{i,j;i+1,j}$$=$$0$, the bond is cut (being absent from 
the graph).  For the square-octagonal graph use of 
Eq.~(\ref{Eq:1BondDragon2}) is appropriate because 
we utilize a ${\vec v}_{\rm Dragon}$ with every element 
equal to $1/\sqrt{m}$ and there is at most one 
intra-slice bond attached to any atom.  

Fig.~\ref{Fig:2D:SqOct}{\bf (A)} shows a nanodevice based on 
a square-octagonal graph with 10\% of the intra-slice bonds cut, chosen in a 
completely random manner.  The remaining intra-slice bonds 
were chosen uniformly to be in $t_{i,j;i+1,j}=[0.8,1.2]$.  
The number of energy values calculated was 142.  
The on site energies were chosen to satisfy Eq.~(\ref{Eq:1BondDragon2}).  
Fig.~\ref{Fig:2D:SqOct}{\bf (B)} shows the underlying graph for {\bf (A)}, 
without leads, with this graph representation more clearly showing 
the cut bond locations.  
For no added disorder, the device is a quantum dragon with 
${\cal T}(E)$$=$$1$, and shows order amidst disorder (not shown) as the 
LDOS$_{i,j}(E)$ is uniform (independent of $i$ and $j$) 
\cite{Novotny_2023}
throughout the nanodevice for any $-2$$<$$E$$<$$2$.  

Fig.~\ref{Fig:2D:SqOct}{\bf (C)} shows both the quantum dragon 
transmission (light cyan horizontal line ${\cal T}(E)=1$), 
as well as the transmission averaged 
over $10^3$ different values of additional 
Gaussian-distributed uncorrelated on~site disorder with disorder strengths 
$\delta=0.02, \> 0.04, \> 0.06$ 
corresponding to red, black, and blue points respectively.  
Fig.~\ref{Fig:2D:SqOct}{\bf (D)} replots the same values for 
the average transmission, but scaled according to Eq.~(\ref{Eq:T-scaling}), 
with the same color coding as in {\bf (C)}.  
Also shown (green) is the DOS$(E)$ for the underlying device shown in 
{\bf (A)}, with the points averaged over the number of eigenvalues of 
the device Hamiltonian ${\cal H}$ using a box-counting method 
with $\Delta E=0.1$.  
Here we have used one energy-independent 
adjustable parameter $L_{\rm scale}\approx 0.7 L$.   
The comparison with our theory of 
Eq.~(\ref{Eq:T-scaling}) is very good.


\section{Conclusion and Discussion}
\label{sec:7:ConcDisc}

We have provided for tight binding Hamiltonians an analysis of the scaling 
upon the addition of uncorrelated disorder 
for nanodevices that have translational order and 
have quasi-ballistic electron transmission, or nanodevices that 
are nearly quantum dragon nanodevices.  
For both ballistic nanodevices and quantum dragon nanodevices the 
electron transmission is ${\cal T}(E)=1$ for all $-2<E<2$ for our 
thin leads. 
In summary, 
we derived the scaling relations as 
a function of $E$ to be 
\begin{equation}
\label{Eq:scale:Sec7:summary}
{\cal T}_{\rm ave}(E) = \!
\left\{
\begin{array}{lclcl}
1 \> - \> \frac{\delta^2}{4-E^2} \> \frac{\ell}{m} 
& \>
& {\rm very\> small\>} \delta 
\\
\>
\\
\frac{1}
{1+2 \pi \delta^2 L_{\rm scale} \frac{{\rm DOS}(E)}{\sqrt{4-E^2}}}
& & {\rm intermediate\>} \delta
\>.
\end{array}
\right.
\end{equation}
We tested via large-scale numerical simulations 
these two scaling predictions, with the analysis with no 
adjustable parameters for 
very small $\delta$ detailed 
in Sec.~\ref{sec:5:SmallDelta} and in Sec.~\ref{sec:6:ScaleDOS}
for intermediate $\delta$ with at most one fitting 
constant $L_{\rm scale}/L$.  
It is important to note that $L_{\rm scale}$ is independent of the energy $E$ 
or the uncorrelated disorder strength $\delta$.  

The derived scaling can be considered 
to use the single parameter scaling in 
$s=L_{\cal T}/L_{\rm scale}$ with $L_{\rm scale}$ proportional to the length $L$ of the nanodevice along the direction of electron flow and $L_{\cal T}$ defined in 
Eq.~(\ref{Eq:L-T}). 
For long quasi-ballistic nanodevices at the 
Fermi energy $E_{\rm F}$ this single parameter scaling was known 
previously \cite{MAR2007} with the scaling parameter $s=L_e/L$ with 
$L_e\approx L_{\cal T}$ the electron 
elastic mean free path.  
This single parameter scaling for ${\cal T}_{\rm ave}$ 
as a function of $L$ 
is also known for 1D ordered systems with small added 
disorder where the scaling goes from being dominated 
in the quasi-ballistic regime by $L_e$ to being dominated in the localization 
regime by $\xi_{\rm A}$ \cite{Dossetti2004}.  

Our study directly impacts different areas, each described in a subsection below.  

\subsection{Analysis of disordered quantum models}

Ordered systems are analyzed theoretically by exploiting translational 
invariance.  For quantum systems this typically involves 
using Bloch wavefunctions and a band structure 
analysis 
\added[id=MAN]{\cite{Buerkle2023}}.
Hence most textbooks on solid state physics start 
with crystal structures and exploit the crystal structure 
to obtain electron transport properties
\cite{Kitt1996,Ashc1976,Hofm2015}.  
For systems with weak disorder, disordered 
models are taken to be due to perturbations about the 
ordered crystalline structure.  
Because of the extensive nature of the disorder in the 
\added[id=MAN]{quantum dragon}
model 
Hamiltonians and graphs we study, 
these traditional theoretical tools cannot be applied.  

For example, consider taking a Fourier transform of 
the Hamiltonian of a nanodevice.  If there is translational 
invariance one then has a Hamiltonian that can be 
written as a sum over Fourier coefficients, and assigns 
the Fourier coefficient with wavevector ${\vec k}$ 
to a quasi-particle electron with that wavevector.  
For a quantum dragon nanodevice Hamiltonian the Fourier transform 
can be formally written.  
However, and in particular for the case of a general ${\vec v}_{\rm Dragon}$, 
the Fourier coefficients will not correspond to 
a useful quasi-particle description.  

Rather than using translational invariance to analyze 
coherent electron transport, 
we have generalized the mapping method \cite{NOVO2014}, 
as detailed in Sec.~\ref{sec:2:ModelMethod} and sketched in Fig.~\ref{Fig:FindDragon}.  
This allows one to calculate the electron transmission 
${\cal T}(E)$ when a 
nanosystem is connected to two semi-infinite uniform leads.  
This method does not always work, in fact it is only for 
configurations with atypical disorder 
that the mapping method is useful.  
In some instances where the mapping method works, 
as sketched in Fig.~\ref{Fig:FindDragon}, 
the mapping yields a uniform 1D wire and the nanodevice 
is a quantum dragon with ${\cal T}(E)=1$ for all electron 
energies $E$ in some finite range.  We have shown 
the quantum dragons exhibit
{\bf order amidst disorder\/}, for 
example the LDOS$_{i,j}(E)$ evince translational 
invariance \cite{Novotny_2023}.  

\subsection{Nearly quantum dragon nanodevice circuits}

No disordered quantum dragon nanodevice or ballistic translationally-invariant 
nanodevice is expected to be perfect with ${\cal T}(E)=1$ for a range of 
energies, due to uncorrelated disorder, edge effects for 
stable nanodevices, and/or operation at finite temperature.  
We have derived and demonstrated two scaling regimes 
for nanodevices which are very close to quantum dragon nanodevices 
or ballistic nanodevices.  
Extremely close to a quantum dragon nanosystem, 
we find a universal scaling, Eq.~(\ref{Eq:deltaSmall:01}), 
that does not depend on any device parameters other 
than $\ell$, $m$, the energy of the incoming electron, 
and variance $\delta^2$ of the uncorrelated site disorder added 
to the quantum dragon nanosystem. 
For larger, but still small, added uncorrelated site disorder 
our scaling of Eq.~(\ref{Eq:T-scaling}) gives 
predictions for the average transmission ${\cal T}_{\rm ave}(E)$ 
for an energy $E$ of the incoming electron.  The prediction 
only depends on the length $L$ along the direction of current flow, 
$\delta^2$, $E$, and the density of states per atom 
DOS$(E)$.  In both scaling cases the expected transmission 
at any energy is very close to unity.  
Therefore, it is very reasonable to assume quantum dragon 
nanodevices can be synthesized, constructed, measured, 
and utilized as parts of electronic, opto-electronic, 
or spintronic circuitry.  For example, just as ballistic 
diodes and transistors utilize the ${\cal T}(E)\approx 1$ property 
for ballistic electron propagation, quantum dragon 
diodes and transistors could utilize the ${\cal T}(E)\approx 1$ property.  

The measurement, understanding, and utilization of ballistic electron 
propagation is over a hundred years old.  For example, it gives 
the current-voltage characteristics of vacuum tubes, where it is 
known as the Child-Langmuir law 
\cite{Chil1911,Lang1913}.  
Carbon based nanotubes have been 
utilized as quantum wires \cite{TANS1997} 
and made into 
transistors, including field-effect transistors (FETs) 
\cite{POS2001,JAV2003,CHE2005,SCH2010,LLI2017}.  
More recently, exploiting the ${\cal T}(E)\approx 1$ property has more been 
used to build a ballistic rectifier fabricated in single-layer graphene 
sandwiched by boron nitride flakes, exhibiting a 
mobility of $\sim 2\times 10^5$~cm$^2$/V$\cdot$s and a voltage 
responsivity of $2.3\times 10^4$~V/W with these properties 
holding to room temperature \cite{Auto2016}.  
The ${\cal T}(E)\approx 1$ property also has allowed ballistic diodes 
operating in the THz range \cite{Koch2019}, due to the way electrons 
with ${\cal T}(E)\approx 1$ and large 
$v(E)$ (as in \ref{CSaF_AppC}) 
interact with electromagnetic radiation.  
The ${\cal T}(E)\approx 1$ \added[id=MAN]{property}
has also been used to design 
a ballistic deflection transistor \cite{Didu2006} and very 
recently a room temperature diode based on a ratcheting technique \cite{Cust2020}.  

Therefore we speculate that just as quasi-ballistic nanodevices can be built using 
the ${\cal T}(E)$$\approx$$1$ property, quasi-quantum dragon nanodevices could 
be designed and built using the ${\cal T}(E)$$\approx$$1$ property.  One of the main 
obstacles of designing a quasi-ballistic device is that minor changes to the geometry 
or disorder causes strong scattering of the electrons and ${\cal T}(E)$ rapidly 
becomes small.  We here demonstrated that quasi-quantum dragon devices can have 
allowed changes to the geometry 
(for example by cutting bonds) 
and {\bf locally\/} in the Hamiltonian 
to the disorder all while keeping the ${\cal T}(E)$$\approx$$1$ 
property.  These allowed local (in the Hamiltonian) and global (in the 
geometry and topology) changes can be incorporated into designs of 
nanodevices based on being nearly quantum dragons.  We anticipate that now 
that we have predicted the ${\cal T}(E)$$\approx$$1$ property for such nanosystems 
they can be made into excellent sensors for electromagnetic radiation and 
magnetic fields as well as into diodes and transistors.  

\added[id=MAN]{It is important to also understand 
the limitations of our study, which has utilized 
the single-band tight-binding model only. 
Future studies should add electron spin into the model, 
as well as interactions with quantized crystal vibrations 
(which are phonons in translationally invariant 
systems).  
The next level of chemical accuracy could utilize DFT,
but even for point defects in graphene DFT is 
insufficient and quantum Monte Carlo should be used 
\cite{Thomas2022PRBqmc}.   
Unfortunately, increasing chemical accuracy through 
either DFT or quantum Monte Carlo decreases the number 
of atoms that can be simulated, making such simulations for 
a device with many hundreds of thousands of atoms 
unrealistic.  
Another option would be to use a
modified embedded atom method with bond order potential (MEAM-BO) to capture mechanical properties
\cite{Ababtin2022}, and couple this with a realistic tight-binding model for electronic transport.  
Such calculations would make quantitative comparison 
with experiments easier, particularly for simpler 
quantum dragon nanosystems such as those in 
Fig.~\ref{Fig:SplitTubes}.  
One could also add dissipative effects, which may allow one to study 
the Liouvillian skin effect in the open quantum systems 
\cite{zhang2022LSEliouvillian,li2022LSEquantum,liu2023LSEexperimental,Begg2024LSE}.  
One synthesis route for quantum dragon nanomaterials may be related to nanoporous graphene \cite{Qin2024nanoporeGraphene}.  
However, perhaps experimentally the most direct path to 
test our predictions 
would be to conduct transport measurements of 
a partially-unzipped carbon nanotube, such as 
is depicted in Fig.~\ref{Fig:SplitTubes}B, as 
then the contact hopping terms can be handled in 
the same way as has been accomplished in studies of electrical  
transport 
in uniform single-wall carbon nanotubes \cite{yao2000high}.  
We anticipate even such experimental 
and more realistic calculations will still exhibit 
the nearly-quantum dragon effect 
with ${\cal T}_{\rm ave}(E)$$\approx$$1$ and scaling 
as we 
have studied herein.
}

\subsection{Extended states in $\mathrm{D}>1$}

One conclusion addresses a long-standing open problem in 
mathematical physics.  The problem 
\added[id=MAN]{in D$>$$1$ 
\cite{Carm1990,Aizenman1993,Stolz2011}
}  
is to
{\it establish the existence (in some energy range) of 
extended eigenstates, or a continuous spectrum, for linear operators 
with {\bf extensive\/} disorder.  
A prototypical example is the discrete 
Schr{\"o}dinger operator with a random potential.}
We have studied the electron transport in the discrete 
Schr{\"o}dinger operator, namely in the tight-binding 
model in Eq.~(\ref{Eq:HamTB}).  
Because the quantum dragons have electron transmission 
${\cal T}(E)$$=$$1$ in the energy range $-2$$<$$E$$<$$2$, there must be at least 
one extended state to allow this transmission.  The extended state 
manifests itself not in the physical basis where one writes the Hamiltonian 
as operators on the sites and bonds of a graph in D dimensions 
(in graph theory the 
sites are called vertices and the bonds are called 
edges \cite{CHAR2012}).  
Rather the extended state manifests itself in the 
\lq rotated' or \lq mapped' \cite{NOVO2014} basis as sketched in 
Fig.~\ref{Fig:FindDragon}.  
The disorder we study is {\bf extensive\/}, in that for example in 
Fig.~\ref{Fig:FindDragon} we could choose the intra-slice hopping 
terms $t_{i,j,i',j}$ from any distribution, and still be able to 
satisfy the general constraint imposed by 
Eq.~(\ref{Eq:FindDragon:A:B}) for assigning the 
on site energies $\epsilon_{i,j}$.  
Eq.~(\ref{Eq:FindDragon:A:B}) imposes only a 
{\bf local constraint\/} on the disorder.  
In 1D the presence of any (locally correlated) disorder leads to 
Anderson localization
\added[id=MAN]{\cite{ANDE1958,LAGE2009,Stolz2011},} 
and hence no extended states.  
We therefore postulate the following theorem:
\begin{itemize}
\item[$\>$] {\bf Theorem:}~ In any dimension $\mathrm{D}>1$ for the 
tight binding Hamiltonian there exists atypical, extensive, 
locally-correlated disordered systems which have at least one 
extended state and have electron transmission equal to 
unity for some energy range. 
\end{itemize}
\added[id=MAN]{Note in contrast to the case of 
uncorrelated large disorder 
where localization occurs \cite{Aizenman1993,Mard2017,Igloi2018,Wortis2023}, our 
correlated disorder may be arbitrarily large and still 
the system for a range of $E$ has 
${\cal T}(E)$$=$$1$.}

We have proven this theorem by explicit construction in 
\added[id=MAN]{this paper in both}
2D \added[id=MAN]{(Sec.~6)} and 3D \added[id=MAN]{(App.~D)}.  
The generalization to graphs with higher embedding 
dimensions is straightforward, 
as the method given by the local constraint of 
Eq.~(\ref{Eq:FindDragon:A:B}) does not 
depend on the embedding dimension and can be 
generalized to any coordination number.  
\added[id=MAN]{The complete proof of the Theorem 
for D$>$$1$ is presented in App.~E.}
In fact, as the graph coordination number $K$ increases, 
by the central limit theorem the on site energy 
that satisfies Eq.~(\ref{Eq:FindDragon:A:B}) will 
approach its average value.  
In other words, 
using braces $\Big\langle\cdots\Big\rangle$ to denote 
the average value, one has
\begin{equation}
\label{Eq:CD:aveVals}
\Big\langle \epsilon_{i,j} \Big\rangle \longrightarrow 
K \Big\langle t_{i,j;i',j} \Big\rangle 
\>.
\end{equation}
\added[id=MAN]{For a hypercubic lattice one has $K$$=$$2$D.}  
Furthermore, the width of the distribution of 
$\epsilon_{i,j}$ becomes narrower as $K$ becomes larger.  
Hence although in 2D quantum dragons are atypical random 
configurations, in graphs with large $K$ one may expect 
the typical random configuration to either be or 
to be near a quantum dragon.

\subsection{Cloaking in coherent quantum systems}

Quantum dragon nanosystems have ${\cal T}(E)=1$ for a 
finite range of energies.  
They occur in any $\mathrm{D}>1$ for select graphs (ordered or 
disordered) and select quantum Hamiltonians, the system 
is the graph + the quantum Hamiltonian.  
Quantum dragon systems may be based either 
on physically motivated systems as in this paper 
or on more general systems \cite{NOVO2014}.  

In either case, quantum dragons exhibit what may be 
called {\bf coherent quantum cloaking\/}.  
Assume the only way to investigate a black-box system with a 
nanodevice is to study the coherent electron transmission, 
i.e.\ measure ${\cal T}(E)$.  
From such a measurement one cannot tell whether there is a 
uniform wire inside the black-box or one or more 
quantum dragon nanodevices.  In other words, by 
adjusting properly the quantum Hamiltonian of a nanodevice 
the fact there is a nanodevice within the black-box is cloaked 
because all incoming electrons in a finite range of energies 
undergo complete electron transmission.  This lack of reflection 
of the incoming coherent electrons means the nanodevice has been 
cloaked from this quantum measurement.  The cloaking can be 
accomplished with proper modifications of the disordered 
graph and/or quantum Hamiltonian.  

\subsection{Quantum dragons in QIP and QC}

The concept of quantum dragons, and the method of devising  
and analyzing them, may have applications in QIP (Quantum 
Information Processing) and QC (Quantum Computing). 
This may include the possibility of forming qubits and quantum gates using nearly ballistic nanodevices \cite{Guo2009,Dragoman2016,Bertoni2000} 
or quantum dragon nanodevices.  
For quantum computing nanodevices the 
quality of the qubit and quantum gates increases as 
${\cal T}_{\rm ave}(E_{\rm F})$ approaches unity.  

There are special-purpose quantum devices which 
can analyze transport in coherent quantum systems 
\cite{Harris2017,Karamlou2022,Yariv2022}.
Although thus far such devices can only 
analyze 1D transport, future generations of related 
special purpose QCs which can handle $\mathrm{D}>1$ are in 
planning stages. They could be used to 
investigate quantum dragon systems, and systems 
that are nearly quantum dragons.  

Gate-based QC proceeds by performing a series of 
$M_{\rm QC}$ unitary operators on an initial qubit state, and 
performs a measurement in the Cbit (classical bit) 
basis after the computation.  The power of 
QC comes from the non-commutative property of the 
unitary operators.  For $N_{\rm QC}$ qubits the operators 
are $2^{N_{\rm QC}}\times 2^{N_{\rm QC}}$.  
The existence of quantum dragons demonstrates 
interesting properties of quantum systems due to 
a set on non-commuting matrices having a 
common eigenvector.  An open question is whether 
requiring all $M_{\rm QC}$ unitary operators to 
have one or more common eigenvectors leads to 
interesting algorithms and applications in 
QC and QIP.


\section*{Acknowledgements}
\added[id=MAN]{Support is acknowledged 
of grant no.~23-05263K of the 
Czech Science Foundation (TN), 
and
partial support from grant US DOE 
DE-SC0024286 (MAN).
}

\appendix

\section{Uniform, Thin, 1D Wire }
\label{CSaF_AppA}

The 2D, 3D, or 2D+3D nanodevice is connected to two thin 1D 
uniform semi-infinite leads. 
Furthermore as sketched in 
Fig.~\ref{Fig:FindDragon}, the quantum dragon nanodevice 
has a uniform quantum wire in the rotated basis of Hilbert space.   
It is this property that enables complete electron transmission even in the 
disordered nanodevice.  The order amidst disorder 
property \cite{Novotny_2023} 
is obvious only after the physical device Hamiltonian 
undergoes a similarity transformation, but is 
of course present in any basis.  
Therefore, the electron transmission of a uniform quantum wire is 
of extreme importance.  

Let $\epsilon_{\rm Lead}$ be the on site energy value for all lead atoms, 
and $t_{\rm Lead}$ be the hopping strength between 
nn pairs of atoms in the uniform wires.  
The right lead has the tight binding Hamiltonian 
\begin{equation}
\label{Eq:WireH}
{\cal H}_{\rm R,Lead} = 
\epsilon_{\rm Lead} \sum\limits_{j=j_0}^\infty  c_j^\dagger c_j - 
t_{\rm Lead} 
\sum\limits_{j=j_0}^\infty  
\left(c_j^\dagger c_{j+1} + c_{j+1}^\dagger c_j\right)
\end{equation}
with $j_0$ the index of the lead atom attached to the nanodevice.  

Let $a$ be the nn distance between lead atoms.  
A Bloch wavefunction analysis for the 
uniform wire gives the dispersion relation 
\begin{equation}
E-\epsilon_{\rm Lead}=-2 t_{\rm Lead} \cos\left(q_{\rm Lead} a\right) 
\end{equation}
with electron wavevector in the leads $q_{\rm Lead}$.   
The lead wires are uniform so they can be regarded as a 
short circuit device, which has complete transmission, 
\begin{equation}  
{\cal T}(E)=1 \> {\rm for \> all \>} 
-2 t_{\rm Lead}< E - \epsilon_{\rm Lead} < 2 t_{\rm Lead}
\>.  
\end{equation}
Thus a uniform wire means complete transmission for all 
electron wavelengths which propagate in the leads, namely 
$2 a\le \lambda_{\rm Lead} <\infty$.  
A transfer matrix method analysis actually shows there is 
complete transmission for 
$-2t_{\rm Lead}\le E\le 2 t_{\rm Lead}$ 
even in a non-uniform case as long as 
$\left\vert t_{{\rm Lead}; \> j,j+1}\right\vert$$=$$1$ for each hopping term 
between nn atoms in 
the 1D wire, i.e.\ phases are allowed as they do not change the 
electron transmission in 1D from being ${\cal T}(E)=1$.  

In this paper we usually choose as the zero of energy 
$\epsilon_{\rm Lead}=0$.  
We also choose as our unit of energy the hopping strength 
$t_{\rm Lead}=1$.  
Hence we have set the zero of energy as well as the units of energy.  
There is a similar Hamiltonian for the left lead, again with 
$\epsilon_{\rm Lead}=0$ and $t_{\rm Lead}=1$.  

We can analyze a length $\ell$ section of the thin wire by 
introducing the thin wire Hamiltonian (written for $\ell$$=$$4$) 
as
\begin{equation}
\label{Eq:AppA:Hwire}
{\cal H}_{\rm wire} 
\> = \> 
\left(\begin{array}{rrrr}
0 & -1 & 0 & 0 \\
-1 & 0 & -1 & 0 \\
0 & -1 & 0 & -1 \\
0 & 0 & -1 & 0 
\end{array}\right)
\end{equation}
leading to the wire Green's function for electrons 
injected with energy $E$
\begin{equation}
\label{Eq:AppA:Gwire}
{\cal G}_{\rm wire}(E) \> = \>  
\Big(
E {\bf I} - {\cal H}_{\rm wire} 
-\mathbf{\Sigma}_{L}(E)
-\mathbf{\Sigma}_{R}(E)
\Big)^{-1}
\>.
\end{equation}
The solution for ${\cal G}_{\rm wire}$ is described in 
\cite{Econ2006}.  
In particular, the matrix elements are given by
\begin{equation}
\label{Eq:AppA:GFelements}
\langle a | {\cal G}_{\rm wire}|b\rangle 
\> = \> 
\frac{-i \> \exp\Big(\> i \> \phi(E) \> |a-b| \Big)}{\gamma(E)}
\end{equation}
for $\ell>1$ 
with 
$\phi(E)={\rm arccos}\left(-\frac{E}{2}\right)$ 
and $\gamma(E)$ defined after Eq.~(\ref{Eq:NEGF:05}).  
Here $a$ and $b$ are the indices for the locations 
along the wire with $1\le a,b\le\ell$. 
\section{Derivation of Scaling for Very Small $\delta$}
\label{CSaF_AppB}

We want to study the lowest perturbative correction to the quantum dragon
or ballistic transmission due to a very weak onsite disorder. 
We start from the 
trace formula Eq.~(\ref{Eq:NEGF:06}) 
for the transmission coefficient in terms of the Green's 
function (GF)
.
The Green's function Eq.~(\ref{Eq:NEGF:04}) is basically
just the resolvent of the system plus leads Hamiltonian projected by
the partitioning method onto the system, i.e.~the quantum dragon subspace. It is
given by 
\begin{equation}
\label{Eq:AppB:02}
\mathcal{G}(E)=
\Big( E {\bf I}_N -\mathcal{H-V}
-\mathbf{\Sigma}_{L}(E)
-\mathbf{\Sigma}_{R}(E)\Big)^{-1}
\end{equation} 
where as in Eq.~(\ref{Eq:NEGF:03}) 
the lead self-energies are 
$\mathbf{\Sigma}_{L}(E)\equiv|L\rangle\sigma(E)\langle L|$
and $\mathbf{\Sigma}_{R}(E)\equiv|R\rangle\sigma(E)\langle R|$ with $\sigma(E)\equiv(E-i\sqrt{4-E^{2}})/2$.  
In Eq.~(\ref{Eq:AppB:02})  
$\mathcal{H}$ is the unperturbed device Hamiltonian, 
for either a quantum dragon or a translationally invariant ballistic nanodevice.  
The diagonal matrix 
$\mathcal{V}\equiv\sum_{\alpha}|\alpha\rangle v_{\alpha}\langle\alpha|$
is the perturbing scattering potential composed of site-uncorrelated
disorder potentials satisfying the conditions (overbar denotes the
impurity averaging) $\overline{v_{\alpha}}=0$ and $\overline{v_{\alpha}v_{\beta}}=\delta^{2}\delta_{\alpha\beta}$.
Using the Dyson equation 
$\mathcal{G}^{-1}(E)=\mathcal{G}_{0}^{-1}(E)-\mathcal{V}$ with $\mathcal{G}_{0}(E)$
the unperturbed GF corresponding to the perfect quantum dragon solution 
and making
its perturbative expansion up to the second order in the disorder
\begin{equation}
\label{Eq:Dyson:start}
\begin{array}{lcll}
\mathcal{G}(E) &\approx & \mathcal{G}_{0}(E) \!& +\>\mathcal{G}_{0}(E)\mathcal{VG}_{0}(E)
\\
& & & +\>\mathcal{G}_{0}(E)\mathcal{VG}_{0}(E)\mathcal{VG}_{0}(E)
\end{array}
\end{equation}
we arrive at the expansion of the transmission 
from Eq.~(\ref{Eq:NEGF:06})
(keeping only the zeroth and second order terms since the first order
term will cancel after the disorder averaging anyway) 
\begin{strip}
\begin{equation}
\begin{array}{lcl}
\mathcal{T}(E)-1 & 
\approx & 
\mathrm{Tr}[\mathbf{\Gamma}_{L}(E)\mathcal{G}_{0}(E)\mathbf{\Gamma}_{R}(E)\mathcal{G}_{0}^{\dagger}(E)]-
1
+\mathrm{Tr}[\mathbf{\Gamma}_{L}(E)\mathcal{G}_{0}(E)\mathcal{VG}_{0}(E)\mathbf{\Gamma}_{R}(E)\mathcal{G}_{0}^{\dagger}(E)\mathcal{VG}_{0}^{\dagger}(E)]\\
 & &
 \quad +\mathrm{Tr}[\mathbf{\Gamma}_{L}(E)\mathcal{G}_{0}(E)\mathcal{VG}_{0}(E)\mathcal{VG}_{0}(E)\mathbf{\Gamma}_{R}(E)\mathcal{G}_{0}^{\dagger}(E)]

 +\mathrm{Tr}[\mathbf{\Gamma}_{L}(E)\mathcal{G}_{0}(E)\mathbf{\Gamma}_{R}(E)\mathcal{G}_{0}^{\dagger}(E)\mathcal{VG}_{0}^{\dagger}(E)\mathcal{VG}_{0}^{\dagger}(E)]\\
 & = & 
\mathrm{Tr}[\mathbf{\Gamma}_{L}(E)
\mathcal{G}_{0}(E)\mathcal{VG}_{0}(E)
\mathbf{\Gamma}_{R}(E)\mathcal{G}_{0}^{\dagger}(E)\mathcal{VG}_{0}^{\dagger}(E)]
+2\Re\Big\{\mathrm{Tr}[\mathbf{\Gamma}_{L}(E)\mathcal{G}_{0}(E)\mathcal{VG}_{0}(E)
\mathcal{VG}_{0}(E)\mathbf{\Gamma}_{R}(E)
\mathcal{G}_{0}^{\dagger}(E)]\Big\}.
\end{array}
\label{eq:trans-expansion}
\end{equation}


Having done this, it's time to use the specifics of quantum dragon or ballistic transmission 
solutions,
in particular the unitary transform $\mathbf{U}_{N}$ to the
\lq\lq uniform wire plus disconnected rest'' basis (represented in 
Fig.~\ref{Fig:FindDragon}). This will help us significantly simplify
the general formula of Eq.~(\ref{eq:trans-expansion}) above. 
In particular,
we use the fact that the transformed coupling matrices have the 
following form 
$\mathbf{U}_{N}\mathbf{\Gamma}_{L}(E)\mathbf{U}_{N}^{\dagger}
=\mathrm{diag[\gamma(E),0,0,\ldots]}$
and 
$\mathbf{U}_{N}\mathbf{\Gamma}_{R}(E)\mathbf{U}_{N}^{\dagger}
=\mathrm{diag[0,\ldots,\gamma(E),0,0,\ldots]}$
with the nonzero entry at the position $\ell(m-1)+1$. 
Furthermore, the transformed unperturbed, 
i.e.~quantum dragon Green functions $\mathcal{G}_{0}(E)$, 
have a block-diagonal form reading 
\begin{equation}
\label{Eq:AppB:UtoGwire}
\mathbf{U}_{N}\mathcal{G}_{0}(E)\mathbf{U}_{N}^{\dagger}=\left(\begin{array}{cc}
\mathcal{G}_{\mathrm{wire}}(E) & 0\\
0 & \mathcal{G}_{\mathrm{rest}}(E)
\end{array}\right)
\end{equation}
with the standard 1-D wire GF having matrix elements 
\cite{Econ2006} given in Eq.~(\ref{Eq:AppA:GFelements}).  
Plugging these transformed quantities
into the first term on right hand side of the first line in Eq.~(\ref{eq:trans-expansion})
we get 
\begin{equation}
\begin{array}{lcl}
\mathrm{Tr}[\mathbf{\Gamma}_{L}(E)\mathcal{G}_{0}(E)\mathbf{\Gamma}_{R}(E)\mathcal{G}_{0}^{\dagger}(E)] 
& \> = \> & 
\gamma^{2}(E)\langle L|\mathcal{G}_{0}(E)|R\rangle\langle R|\mathcal{G}_{0}^{\dagger}(E)|L\rangle\\
& = & \gamma^{2}(E)\langle1|\mathcal{G}_{\mathrm{wire}}(E)|\ell
\rangle\langle \ell |\mathcal{G}_{\mathrm{wire}}^{\dagger}(E)|1\rangle\\
& = & 
\gamma^{2}(E)|\langle1|\mathcal{G}_{\mathrm{wire}}(E)|\ell\rangle|^{2} \\
& = & 1
\end{array}
\label{eq:unitary-transmission}
\end{equation}
which is just the quantum dragon solution, 
or for translationally invariant systems the ballistic electron 
propagation, as it must be. 

Now, let's consider the first term of the last line in Eq.~(\ref{eq:trans-expansion})
and perform the disorder averaging. We get the following expression
for this term denoted as ``ver'' meaning \emph{vertex correction}
\begin{equation}
\begin{array}{lcl}
\overline{\delta\mathcal{T}_{\mathrm{ver}}}  
& \> =\> 
& \mathrm{Tr}[\mathbf{\Gamma}_{L}(E)\mathcal{G}_{0}(E)\overline{\mathcal{VG}_{0}(E)\mathbf{\Gamma}_{R}(E)\mathcal{G}_{0}^{\dagger}(E)\mathcal{V}}\mathcal{G}_{0}^{\dagger}(E)]
\\
& = & \delta^{2}
\sum_{\alpha}\mathrm{Tr}[\mathbf{\Gamma}_{L}(E)\mathcal{G}_{0}(E)|\alpha\rangle\langle\alpha|\mathcal{G}_{0}(E)\mathbf{\Gamma}_{R}(E)\mathcal{G}_{0}^{\dagger}(E)|\alpha\rangle\langle\alpha|\mathcal{G}_{0}^{\dagger}(E)]\\
& = &
\delta^{2}\sum_{\alpha}\langle\alpha|\mathcal{G}_{0}(E)\mathbf{\Gamma}_{R}(E)\mathcal{G}_{0}^{\dagger}(E)|\alpha\rangle\langle\alpha|\mathcal{G}_{0}^{\dagger}(E)\mathbf{\Gamma}_{L}(E)\mathcal{G}_{0}(E)|\alpha\rangle.
\end{array}
\label{eq:deltaTver}
\end{equation}
Here, we prove that the two constituents of the product 
$\langle\alpha|\mathcal{G}_{0}(E)\mathbf{\Gamma}_{R}(E)\mathcal{G}_{0}^{\dagger}(E)|\alpha\rangle$
and $\langle\alpha|\mathcal{G}_{0}(E)\mathbf{\Gamma}_{L}(E)\mathcal{G}_{0}^{\dagger}(E)|\alpha\rangle$
(complex-conjugated in Eq.~(\ref{eq:deltaTver})) 
are actually independent of the site index $\alpha$. 
Using the unitary transform $\mathbf{U}_{N}$ we can 
rewrite the term 
$\langle\alpha|\mathcal{G}_{0}(E)\mathbf{\Gamma}_{L}(E)\mathcal{G}_{0}^{\dagger}(E)|\alpha\rangle$ 
(the other one follows analogously) as $\gamma(E)|\langle1|\mathcal{G}_{\mathrm{wire}}(E)\mathbf{U}_{N}|\alpha\rangle|^{2}$. 
Using Eq.~(\ref{Eq:AppB:UtoGwire}) 
we have 
\begin{equation}
\label{Eq:AppE:MAN1}
\begin{array}{lcl}
\langle1|\mathcal{G}_{\mathrm{wire}}(E)\mathbf{U}_{N}|\alpha\rangle
& \> = \> & 
\sum\limits_{a=1}^{\ell}\langle1|\mathcal{G}_{\mathrm{wire}}(E)|a\rangle(\mathbf{U}_{N})_{a,\alpha}
\\
& = & \langle1|\mathcal{G}_{\mathrm{wire}}(E)|j\rangle/\sqrt{m}
\end{array}
\end{equation}
so that altogether
\begin{equation}\label{eq:identity}
\langle\alpha|\mathcal{G}_{0}(E)\mathbf{\Gamma}_{L}(E)\mathcal{G}_{0}^{\dagger}(E)|\alpha\rangle
=\frac{1}{m\>\gamma(E)},
\end{equation}
which is manifestly independent of the site index $\alpha$. 
The right term leads to the very same result so that we finally get
\begin{equation}
\begin{array}{lcl}
\overline{\delta\mathcal{T}_{\mathrm{ver}}} &
\> = \> & 
\delta^{2}\sum\limits_{\alpha}\langle\alpha|\mathcal{G}_{0}(E)\mathbf{\Gamma}_{R}(E)\mathcal{G}_{0}^{\dagger}(E)|\alpha\rangle
\langle\alpha|\mathcal{G}_{0}^{\dagger}(E)\mathbf{\Gamma}_{L}(E)\mathcal{G}_{0}(E)|\alpha\rangle\\
& = &
\delta^{2}\sum\limits_{\alpha}\left(
\frac{1}{m\gamma(E)}\right)^{2}
\\
& = & \frac{\delta^{2}\>\ell}{m\gamma^{2}(E)}.
\end{array}
\label{eq:deltaTver-1}
\end{equation}

Let's now address the second term in the last line of Eq.~\ref{eq:trans-expansion}
corresponding to the renormalization of the single-particle GF ---
performing the disorder averaging we obtain (note that latin indices
such as $a,\, b$ denote the wire subspace spanned by sites $1\ldots \ell$
while the greek ones such as $\alpha$ are site indices in the original
tight-binding basis for the whole dragon structure, 
i.e.~$\alpha=\{i, j\}$ with $i=1\dots m,\,j=1\dots \ell$)
\begin{equation}
\begin{array}{lcl}
\overline{\delta\mathcal{T}_{\mathrm{ren}}} 
& \> = \> &
2\Re\{\mathrm{Tr}[\mathbf{\Gamma}_{L}(E)\mathcal{G}_{0}(E)\overline{\mathcal{VG}_{0}(E)\mathcal{V}}\mathcal{G}_{0}(E)\mathbf{\Gamma}_{R}(E)\mathcal{G}_{0}^{\dagger}(E)]\}
\\
& = & 
2\Re\{\mathrm{Tr}[\gamma(E)|L\rangle\langle L|\mathcal{G}_{0}(E)\overline{\mathcal{VG}_{0}(E)\mathcal{V}}\mathcal{G}_{0}(E)\gamma(E)|R\rangle\langle R|\mathcal{G}_{0}^{\dagger}(E)]\}\\
& = & 
2\delta^{2}\Re\{\sum\limits_{\alpha}\langle L|\mathcal{G}_{0}(E)|\alpha\rangle\langle\alpha|\mathcal{G}_{0}(E)|\alpha\rangle\langle\alpha|\mathcal{G}_{0}(E)|R\rangle\gamma^{2}(E)\langle R|\mathcal{G}_{0}^{\dagger}(E)|L\rangle\}\\
& = &
2\delta^{2}\Re\left\{ \frac{\sum_{\alpha}\langle L|\mathcal{G}_{0}(E)|\alpha\rangle\langle\alpha|\mathcal{G}_{0}(E)|\alpha\rangle\langle\alpha|\mathcal{G}_{0}(E)|R\rangle}{\langle L|\mathcal{G}_{0}(E)|R\rangle}\right\} 
\\
& = & 
2\delta^{2}\Re\left\{ \mathcal{G}_{0}^{\mathrm{loc}}(E)\frac{\sum_{\alpha}\langle L|\mathcal{G}_{0}(E)|\alpha\rangle\langle\alpha|\mathcal{G}_{0}(E)|R\rangle}{\langle L|\mathcal{G}_{0}(E)|R\rangle}\right\} \\ 
& = & 
2\delta^{2}\Re\left\{ \mathcal{G}_{0}^{\mathrm{loc}}(E)\frac{\langle1|\mathcal{G}_{\mathrm{wire}}^{2}(E)|\ell\rangle}{\langle1|\mathcal{G}_{\mathrm{wire}}(E)|\ell\rangle}\right\} 
\\
& = & 
\frac{2\delta^{2}}{\gamma(E)}\Im\left\{ \mathcal{G}_{0}^{\mathrm{loc}}(E)\frac{\sum_{a=1}^{\ell}e^{i\phi(E)(\ell-a)}e^{i\phi(E)(a-1)}}{e^{i\phi(E)(\ell-1)}}\right\}\\
& = & 
\frac{2\delta^{2}\>\ell}{\gamma(E)}\Im\left\{ \mathcal{G}_{0}^{\mathrm{loc}}(E)\right\} .
\end{array}
\label{eq:deltaTren}
\end{equation}
We have used the unitary transmission condition 
of Eq.~(\ref{eq:unitary-transmission})
when going from the second to the third line and the observed fact
that (the imaginary part of) the site-local Green function $\langle\alpha|\mathcal{G}_{0}(E)|\alpha\rangle$ 
from Eq.~(\ref{Eq:AppA:Gwire})
is a site-independent function of energy denoted here by $\mathcal{G}_{0}^{\mathrm{loc}}(E)$.
Now, let's evaluate this quantity 
\begin{equation}\label{eq:loc-GF}
\begin{array}{lcl}
\Im\left\{ \mathcal{G}_{0}^{\mathrm{loc}}(E)\right\}  
& \> = \> & 
\Im\langle\alpha|\mathcal{G}_{0}(E)|\alpha\rangle \\
& = & \frac{1}{2i}\langle\alpha|\mathcal{G}_{0}(E)-\mathcal{G}_{0}^{\dagger}(E)|\alpha\rangle 
\\
& = & 
-\frac{1}{2i}\langle\alpha|\mathcal{G}_{0}(E)[\mathcal{G}_{0}^{-1}(E)-(\mathcal{G}_{0}^{\dagger}(E))^{-1}]\mathcal{G}_{0}^{\dagger}(E)|\alpha\rangle\\
 & = &
 -\frac{1}{2}\langle\alpha|\mathcal{G}_{0}(E)[\mathbf{\Gamma}_{L}(E)+\mathbf{\Gamma}_{R}(E)]\mathcal{G}_{0}^{\dagger}(E)|\alpha\rangle
\\
 & = & -\frac{1}{\gamma(E)\>m},
\end{array}
\end{equation}
\end{strip}
$\!\!\!\!$ where we have used 
Eq.~(\ref{eq:identity}) 
in the last step. 
This finally leads to the expression 
\begin{equation}
\overline{\delta\mathcal{T}_{\mathrm{ren}}}
\> = \> -\frac{2\>\delta^{2}\>\ell}{\gamma^{2}(E)\>m},
\end{equation}
and altogether we arrive at the final small~$\delta$ universal
scaling law reading
\begin{equation}
\label{Eq:AppB:final}
\begin{array}{lcl}
\mathcal{T}_{\mathrm{ave}}(E) 
& \>=\> & 
1-\frac{\delta^{2}}{4-E^{2}}\>\frac{\ell}{m} 
\>. \\
\end{array}
\end{equation}

For a uniform ${\vec v}_{\rm Dragon}$ but with varying values 
of the number of atoms in the $j^{\rm th}$ slice, $m_j$, 
the scaling uses a 
value $m_{\rm Scale}$ for $m$, with 
$m_{\rm Scale}$ given by \cite{Novotny_2021} 
\begin{equation}
\label{Eq:mScale}
\frac{1}{m_{\rm Scale}} \> = \> 
\frac{1}{\ell} \>
\sum\limits_{j=1}^\ell \frac{1}{m_j}
\>.
\end{equation}
In Eq.~(\ref{Eq:AppB:final}) it is important to keep in mind 
the range of validity for the scaling.  
For example, although for large $m$ one has 
${\cal T}(E)\rightarrow 1$, this result is valid only close to 
at most one Fano resonance.  As $m$ or $\ell$ increase for a given 
type of Hamiltonian the density of the Fano resonance singularities, 
the DOS$(E)$, also increases so the validity of 
Eq.~(\ref{Eq:AppB:final}) is then valid only for smaller 
intervals of energy.  

Furthermore, using this method with the unitary matrices 
we have also explicitly evaluated the local 
density of states (LDOS) for the propagating 
electrons.  The result using Eq.~(\ref{Eq:NEGF:LDOS}) is 
\begin{equation}\label{eq:LDOSapp}
\begin{array}{lcl}
\mathrm{LDOS}_{i,j}(E) & \> \equiv \> &  -\frac{1}{\pi}\Im\langle\alpha|\mathcal{G}_{0}(E)|\alpha\rangle
\\
\\
&= & \frac{1}{\pi \> m_j \> \sqrt{4-E^{2}}}. \\
\end{array}
\end{equation}
In \cite{Novotny_2023,Novotny_2021} we showed the LDOS for 
quantum dragon nanodevices may exhibit 
order-amidst-disorder.  
For example, for nanodevices with uniform $m$ and our ${\vec v}_{\rm Dragon}$ 
the LDOS is the same at every site.  


\section{Derivation of Scaling Related to DOS}
\label{CSaF_AppC}

With increasing disorder strength $\delta$ the situation dramatically 
changes from that of \ref{CSaF_AppB}, 
and the average transmission becomes strongly energy dependent. 
This can be seen in Fig.~\ref{Fig:DragonFewFano}.  
This corresponds to the regime of a number of nearby Fano anti-resonances 
being important in calculating ${\cal T}_{\rm ave}(E)$ for a 
specific energy.  
This scaling regime can be described by a generalization of the 
scaling approach of Ref.~\cite{MAR2007} used for 
an analysis of weakly doped long silicon nanowires 
that have nearly ballistic electron transmission.  
Their analysis assumes the linear increase 
of the sample resistance with its length $L$, i.e., 
$\frac{1}{\mathcal{T}_{\rm ave}(E)}=1+\frac{L_{\rm scale}}{L_{e}(E)}$, with $L_{\rm scale}\propto L$ via an unknown, device-dependent but energy and $\delta$-independent constant.  
The mean free path 
for their system is 
$L_{e}(\epsilon_{\mathbf{k}})=v(\epsilon_{\mathbf{k}})
\tau(\epsilon_{\mathbf{k}})$,
where the ballistic velocity of a propagating mode $\mathbf{k}$ 
(relevant for the transmission and, thus, the conductance/resistance) 
$v(\epsilon_{\mathbf{k}})=\sqrt{4-\epsilon_{\mathbf{k}}^{2}}$.  
For our analysis 
the scattering time $\tau(\epsilon_{\mathbf{k}})$ is estimated
from the Fermi golden rule for the scattering potential $\mathcal{V}=\sum_{\alpha}|\alpha\rangle v_{\alpha}\langle\alpha|$
with uncorrelated on site disorder potentials satisfying the condition
(overbar denotes the impurity averaging) $\overline{v_{\alpha}}=0$ and $\overline{v_{\alpha}v_{\beta}}=\delta^{2}\delta_{\alpha\beta}$.   
For our quantum dragon nanodevices, 
one has to be careful about using this formalism. 
For the case of nearly ballistic electron transmission of Ref.~\cite{MAR2007} 
the only scattering is due to the impurities, and because the system is 
almost pure the quantities $\epsilon_{\mathbf{k}}$ and 
$\tau(\epsilon_{\mathbf k})$, and hence $L_e(E)$ make sense in physical space.  
For a disordered quantum dragon the situation is more complicated to interpret.  
Only in the rotated basis, as depicted in Fig.~\ref{Fig:FindDragon}, and for some degrees of 
freedom is 
a quasi-particle interpretation of $\epsilon_{\mathbf{k}}$ sensible, and hence 
only in this not-physical-space basis can the quantities  
$v(\epsilon_{\mathbf{k}})$ and $\tau(\epsilon_{\mathbf k})$ be easily utilized.  
With this caveat, one can use the length of 
Eq.~(\ref{Eq:L-T}) with $L_{\cal T}\approx L_e$.  
The calculation is 
\begin{equation}
\begin{array}{lcl}
\frac{1}{\tau(\epsilon_{\mathbf{k}})} & \> = \> &  
2\pi\sum_{f\mathbf{\neq k}}\overline{\left|\left\langle \mathbf{k}|\mathcal{V}|f\right\rangle \right|^{2}}\delta(\epsilon_{\mathbf{k}}-\epsilon_{f})\\
 & = & 2\pi\left\langle \mathbf{k}|\overline{\mathcal{V}\delta(\epsilon_{\mathbf{k}}
 -\mathcal{H})\mathcal{QV}}|\mathbf{k}\right\rangle \\
 & = &2\pi\sum_{\alpha,\beta}\overline{v_{\alpha}v_{\beta}}\left\langle \mathbf{k}|\alpha\right\rangle \left\langle \beta|\mathbf{k}\right\rangle \left\langle \alpha|\,\delta(\epsilon_{\mathbf{k}}-\mathcal{H})\mathcal{Q}|\beta\right\rangle \\
 & = & 2\pi\delta^{2}\sum_{\alpha}\left|\left\langle \mathbf{k}|\alpha\right\rangle \right|^{2}\left\langle\alpha|\,\delta(\epsilon_{\mathbf{k}}
 -\mathcal{H})\mathcal{Q}|\alpha\right\rangle \\
 & = & 2\pi\delta^{2}\frac{1}{N}\sum_{\alpha}\left\langle\alpha|\,\delta(\epsilon_{\mathbf{k}}
 -\mathcal{H})\mathcal{Q}|\alpha\right\rangle \\
 & \approx & 2\pi\delta^{2}\>\mathrm{DOS}(\epsilon_{\mathbf{k}}),
\end{array}
\end{equation}
where the projector $\mathcal{Q}\equiv\mathrm{1}-|\mathbf{k}\rangle\langle\mathbf{k}|$
and the DOS (per site) is defined as $\mathrm{DOS}(\epsilon_{\mathbf{k}})\equiv\frac{1}{N}\sum_{\alpha}\left\langle \alpha|\,\delta(\epsilon_{\mathbf{k}}-\mathcal{H})|\alpha\right\rangle $.
Neglecting the projector in the transition in the last line we have
just neglected one out of $N$ (eigen)states, which should be a negligible
error on the order of $1/N$. 
We have used the identity 
$\left|\left\langle \mathbf{k}|\alpha\right\rangle \right|^{2}=1/N$, 
which can be proven easily using the rotation matrix 
$\mathbf{U}_{N}$ inserted into the scalar product 
$\langle \mathbf{k}|\alpha\rangle=\langle \mathbf{k}|\mathbf{U}_{N}\mathbf{U}_{N}^{\dagger}|\alpha\rangle$. 
Using the fact that the rotated propagating mode corresponds to an eigenstate of the uniform wire in the transformed basis and 
Eq.~(\ref{eq:similarity-matrix}) 
we get 
\begin{equation}
\begin{array}{lcl}
\langle \mathbf{k}|\mathbf{U}_{N}\mathbf{U}_{N}^{\dagger}|\alpha\rangle 
& = & \sum_{a=1}^{l}\langle \mathbf{k}|\mathbf{U}_{N}|a\rangle\langle a|\mathbf{U}_{N}^{\dagger}|\alpha\rangle 
\\
& = & 
\langle \mathbf{k}|\mathbf{U}_{N}|j\rangle/\sqrt{m} \\
& = & \langle\mathbf{k}_{\mathrm{wire}}|j\rangle/\sqrt{m}=e^{ikj}/\sqrt{lm} \>. 
\end{array} 
\end{equation}
Putting everything together we arrive at the following expression for the transmission in this regime
\begin{equation}\label{Eq:AppC:T-scaling}
\begin{array}{lcl}
\mathcal{T}_{\rm ave}(E) & \>=\> & 
\frac{1}{1+2\pi\delta^{2}L_{\rm scale}\frac{\mathrm{DOS}(E)}{\sqrt{4-E^{2}}}} 
\>.
\end{array}
\end{equation}
This scaling relation is tested on different 2D nanodevices in 
Sec.~\ref{sec:6:ScaleDOS}.

\section{\label{sec:AppD} Quantum Dragons in 3D and 2D+3D}
\label{CSaF_AppD}
In this appendix we describe two types of quantum dragons, both 
are used to test the scaling for very small $\delta$ in 
Sec.~5.  The first subsection is for a 3D disordered nanodevice, 
the second is for a 2D+3D nanodevice.  
Both constructions have a constant number of atoms 
$m$ in each slice.  

\subsection{3D Quantum Dragon}
\label{CSaF_AppD1}

\begin{center}
\begin{figure}[tb]
\includegraphics[width=0.47\textwidth]{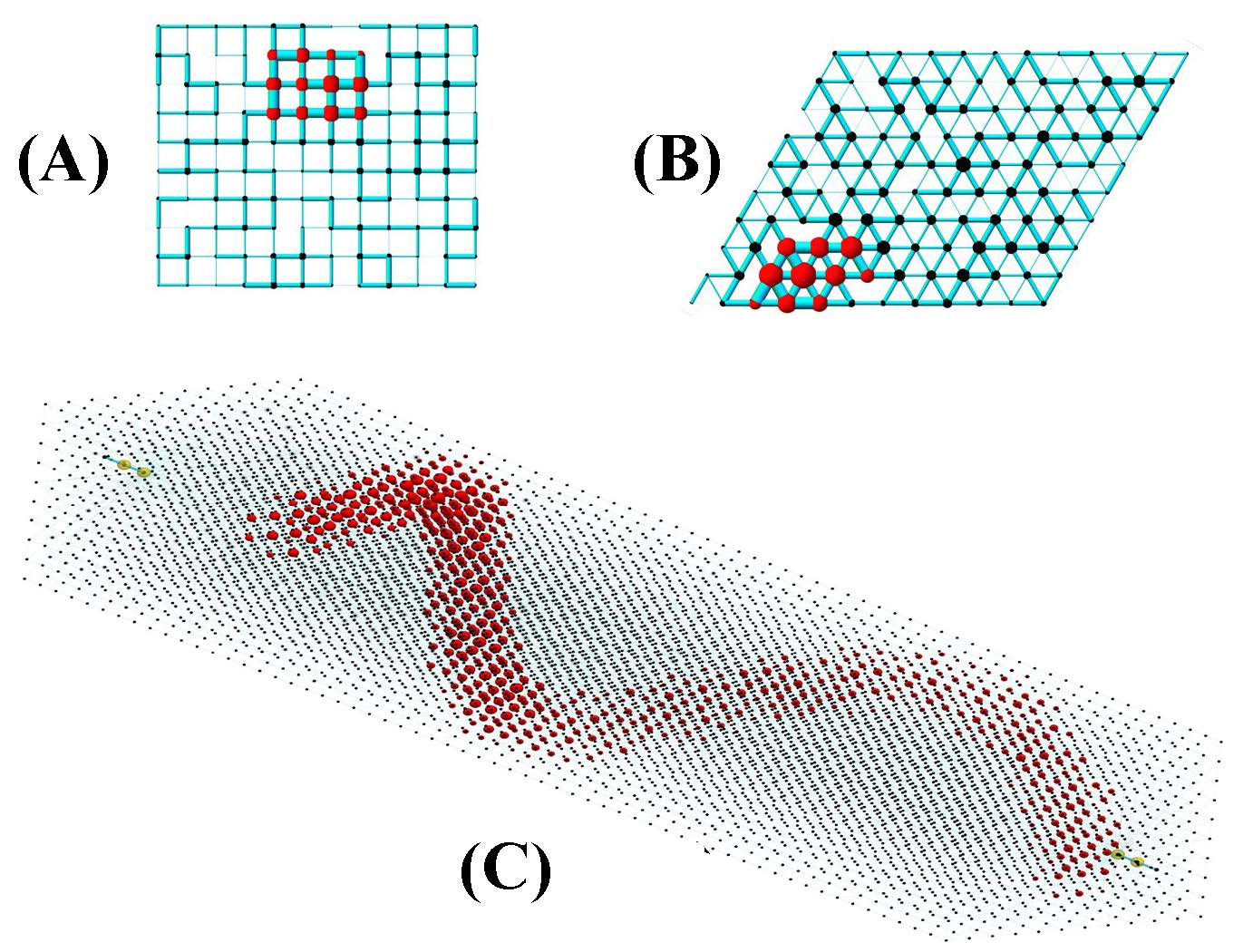}
\caption{
\label{Fig:3D:snake}
Strongly disordered 3D structure with 
$\ell=60$ and $m=120$ based on each 2D slice being $10\times 12$.  
The radii of the two colors of spheres (red and black) are proportional to the on site energies, 
but are for regions 
with different distributions of randomly choosing the nn intra-slice bond strengths.  
All bonds are shown as cyan cylinders, with radii proportional to the 
hopping strength.  
{\bf (A)}~Shows a 2D rectangular graph based slice.  
{\bf (B)}~Shows a 2D triangular graph based slice.  
The entire device is shown in {\bf (C)} with $\ell/2$ 
2D rectangular (left end of device) 
and $\ell/2$ 2D triangular (right end of device) intra-slice sub-graph slices.  
See text in \ref{CSaF_AppD1}\ for a full description.  
}
\end{figure}
\end{center}

It is possible to construct a quantum dragon 
using 2D intra-slice graphs with strong locally-correlated disorder
for each slice.  This gives a 3D overall graph after 
including the inter-slice bonds, 
and still have order amidst disorder and ${\cal T}(E)=1$.  
For a given ${\vec v}_{\rm Dragon}$ with all positive elements, 
and indexing the intra-slice atoms by $i$ 
(even though the intra-slice graph has $m$ atoms arranged on a 2D graph) 
for any random choice of intra-slice bonds 
a quantum dragon nanodevice need to have all on site energies satisfy 
Eq.~(\ref{Eq:FindDragon:A:B}).  
Hence one has also in 3D atypical disorder, but only locally correlated disorder, 
that allows for nanodevices to be quantum dragons and to 
exhibit order amidst disorder.  

An example for a uniform ${\vec v}_{\rm Dragon}$ is shown in 
Figure~\ref{Fig:3D:snake} for $\ell=60$ and $m=120$. In the figure we have chosen the 
inter-slice hopping matrices to all be ${\bf B}_{i,j;i,j+1}=-{\bf I}_{120}$.  
Each slice was considered to be divided into two regions.  
The intra-slice bonds were chosen from three distributions, depending on which region the 
atoms of the bonds belonged to, or if they were bonds between the two regions.   
We restricted ourselves to nn bonds, but had $\ell/2$ slices with a 2D square graph 
(a maximum of 4 nn intra-slice bonds per atom) and 
$\ell/2$ slices with a 2D triangular graph (a maximum of 6 nn intra-slice bonds per atom).  
In region~\#1 (\#2) the nn intra-slice bond strengths were chosen uniformly from the distribution $\left[0,0.5\right]$ 
($\left[0.5,1\right]$).  
Intra-slice nn bonds joining atoms from the two different regions were 
chosen with a 50\% chance to be zero and 
a 50\% chance to have the strength $0.5$.  
The on site energies were chosen to satisfy 
Eq.~(\ref{Eq:FindDragon:A:B}), 
so the nanodevice is 
a quantum dragon.  
Figure~\ref{Fig:3D:snake}{\bf (A)} shows one slice where the underlying 2D graph is a 
rectangular graph with free boundary conditions.
Figure~\ref{Fig:3D:snake}{\bf (B)} shows one slice where the underlying 2D graph is a 
triangular graph with free boundary conditions.
The strongly disordered quantum dragon nanodevice is shown in 
Figure~\ref{Fig:3D:snake}{\bf (C)}.  
The device has ${\cal T}(E)=1$ and exhibits 
order amidst disorder 
\cite{Novotny_2023}
since the LDOS$_{i,j}$ is 
the same for every site.  

\subsection{2D+3D Quantum Dragon}
\label{CSaF_AppD2}

\begin{center}
\begin{figure}[tb]
\includegraphics[width=0.45\textwidth]{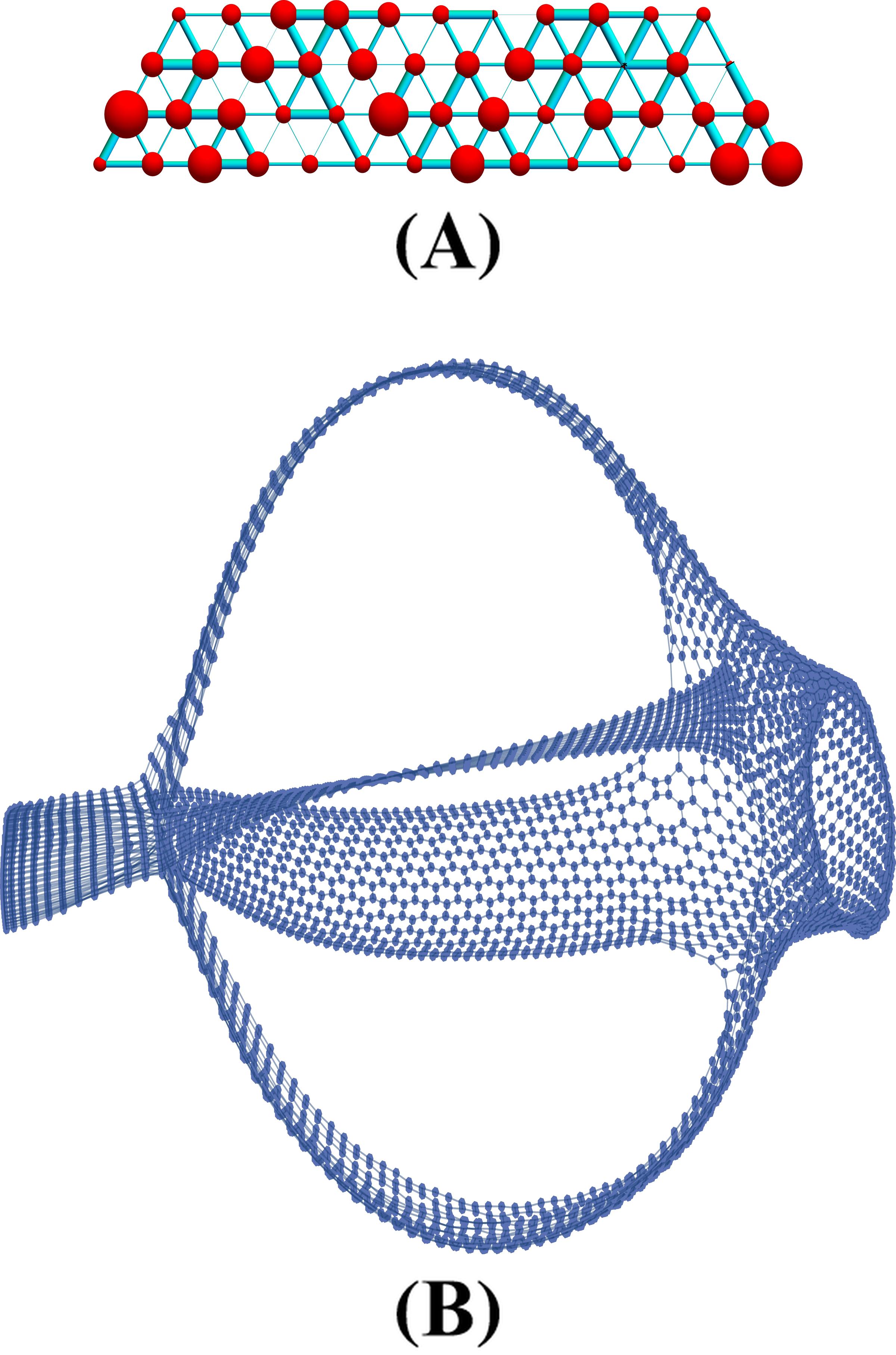}
\caption{
\label{Fig:2D3DmConstant}.  
A combination 2D+3D quantum dragon nanodevice with 
$\ell=82$ and $m=50$.  
{\bf (A)}~Shows a single slice of the 3D hexagonal graph portion of the nanodevice.
The 3D sub-graph has 12 slices. 
{\bf (B)}~Shows the entire device, with the 3D sub-graph 
to the left.  
See \ref{CSaF_AppD1} for further information.  
}
\end{figure}
\end{center}

It is possible to have different embedding dimensions for 
different slices.  
An example of a 2D+3D quantum dragon device is shown in 
Fig.~\ref{Fig:2D3DmConstant}{\bf (B)}, with 
$m=50$ and 
$\ell=82=\ell_{\rm 3D}+\ell_{\rm ribbons}+\ell_{\rm tube}$.  
For convenience all 
inter-slice bonds were set to have $t_{i,j;i,j+1}$$=$$1$, and 
the elements of ${\vec v}_{\rm Dragon}$ were chosen 
to all be equal to $m^{-1/2}$.  

In Fig.~\ref{Fig:2D3DmConstant} 
the 3D sub-graph is composed of $\ell_{\rm 3D}=12$ slices 
of a 2D triangular graph composed of 4 layers with 
$m=50=m_1+m_2+m_3+m_4$ with $m_k$ atoms in layer $k$.  
The graph uses $m_1=11$ and $m_{k+1}=m_k+1$.  The 
intra-slice bonds for the 2D triangular sub-graphs were chosen 
at random with $t_{i,j;i',j}\in[0,1]$.  
If an atom had $K_{i,j}$ intra-slice bonds labeled by $\alpha$ 
with values 
$t_{i,j;i+\alpha,j}$, the on site energies were chosen 
to satisfy 
\begin{equation}
\label{Eq:2D3D:fixed_m}
\epsilon_{i,j} = \sum\limits_{\alpha=1}^{K_{i,j}} t_{i,j;i+\alpha,j}
\end{equation}
in accordance with Eq.~(\ref{Eq:FindDragon:A:B}).  
One of the $\ell_{\rm 3D}$ slices 
of the device in Fig.~\ref{Fig:2D3DmConstant}{\bf (B)}
is shown in 
Fig.~\ref{Fig:2D3DmConstant}{\bf (A)}.  
The radii of the cyan cylinders (red spheres) are 
proportional to the tight binding parameter values for the 
intra-slice hopping terms $t_{i,j;i+\alpha,j}$ (on site 
energies $\epsilon_{i,j}$). The 3D hexagonal sub-graph with 
$\ell_{3D}=12$ is on the left 
in Fig.~\ref{Fig:2D3DmConstant}{\bf (B)}.  

The middle sub-graphs of Fig.~\ref{Fig:2D3DmConstant} 
are composed of ribbons, 
here $\ell_{\rm ribbons}=50$, with 
the same widths $m_k$ as in the four layers of the 
3D sub-graph.  
The intra-slice bonds present in the 2D graph of the 
ribbons were randomly chosen to have strength 
$t_{i,j;i+1,j}\in[0.98,1.02]$, and the 
on site energies satisfy Eq.~(\ref{Eq:2D3D:fixed_m}).  
If an atom has zero intra-slice bonds its 
on size energy is zero.  
The ribbon with $m_1=11$ is a 2D rectangular 
graph, and is given a half-twist before connecting to the 
right-most sub-graph.  
The ribbon with $m_2=12$ is a 2D square-octagonal graph, 
and is given a half-twist before connecting to the 
right-most sub-graph.  
The ribbon with $m_3=13$ is a zigzag 2D hexagonal 
nanoribbon.  
The ribbon with $m_4=14$ is a 2D hexagonal graph, 
with the ends a zigzag ribbon while in the 
middle the ribbon is stitched into a structure 
isomorphic to that of an armchair SWCNT.  

The right-most sub-graph in 
Fig.~\ref{Fig:2D3DmConstant}{\bf (B)} has $\ell_{\rm tube}=20$ 
slices, and the graph is isomorphic to an armchair SWCNT.  
The intra-slice hopping terms were randomly chosen 
to satisfy $t_{i,j;i+1,j}\in[0.9,1.1]$, and the 
on site energies were chosen to satisfy 
Eq.~(\ref{Eq:2D3D:fixed_m}). 

From the paragraphs describing the makeup of the 
graph and tight binding parameters in 
Fig.~\ref{Fig:2D3DmConstant} the amount of disorder 
is evident.  Furthermore, all intra-slice 
hopping parameters were chosen randomly, while the 
disorder for the on site energies are only 
{\it locally correlated\/} since they satisfy 
Eq.~(\ref{Eq:2D3D:fixed_m}).   Nevertheless, the 
nanodevice shown in Fig.~\ref{Fig:2D3DmConstant} 
is a quantum dragon since ${\cal T}(E)=1$ for all 
$-2<E<2$, and for any energy every atom has the 
same value for LDOS$_{i,j}$ (not shown) so the 
device shows order amidst disorder \cite{Novotny_2023}.  

\section{Proof of Sec.~7.3 Theorem.}
\label{CSaF_AppAdded}
We prove the theorem of Sec.~7.3, namely that in any 
dimension $D$$>$$1$ there exists 
locally-correlated disordered tight binding models which are quantum dragons in that they 
have ${\cal T}(E)$$=$$1$ for all 
$-2<E<2$ in our units.  
In particular this requires that we satisfy the two matrix equations of Eq.~\eqref{Eq:FindDragon:A:B}. 
We are free to choose all ${\bf B}_{j,j+1}=-{\bf I}$ with 
${\bf I}$ the identity matrix, thereby satisfying the second 
matrix equation of Eq.~\eqref{Eq:FindDragon:A:B} for the 
inter-slice hopping.  This choice for 
the ${\bf B}_{j,j+1}$ is sufficient, but 
not necessary, for the proof of the theorem.  
We also choose all elements of ${\vec v}_{\rm dragon}$ to be non-zero.

The embedding dimension of the graphs associated with each intra-slice part of the Hamiltonian 
${\bf A}_j$ is $D-1$.  
For each intra-slice hopping 
write ${\bf A}_j={\bf d}_j+{\bf h}_j$ with 
${\bf d}_j$ a diagonal matrix containing the 
on site energy terms and ${\bf h}_j$ containing the 
intra-slice hopping terms. 
The intra-slice portion of 
the Hamiltonian matrix of Eq.~\eqref{Eq:FindDragon:A:B} 
can be written as 
\begin{equation}
\label{Eq:Proof:01}
\left({\bf d}_j+{\bf h}_j\right) {\vec v}_{\rm dragon} = {\vec 0} 
\>.
\end{equation}
Rearranging terms gives the equation 
\begin{equation}
\label{Eq:Proof:02}
{\bf d}_j {\vec v}_{\rm dragon} = 
-{\bf h}_j {\vec v}_{\rm dragon}
\>.
\end{equation}
Thus one locally correlates each on site energy in 
a way that depends only on the $i^{\rm th}$ component of 
the dragon vector, ${\vec v}_{{\rm dragon},i}$, and the 
hopping terms associated with site $i,j$.  
As a reminder, we have chosen all elements of 
${\vec v}_{\rm dragon}$ to be non-zero.  
In particular, 
every on site energy is chosen to be correlated 
with its associated hopping terms as 
\begin{equation}
\label{Eq:Proof:03}
\epsilon_{i,j} = 
\frac{-\left({\bf h}_j {\vec v}_{\rm dragon}\right)_i}
{{\vec v}_{{\rm dragon},i}}
\>.
\end{equation}
\added[id=MAN]{
For a uniform ${\vec v}_{\rm dragon}$ as in this paper, 
Eq.~(\ref{Eq:Proof:03}) reduces to Eq.~(\ref{Eq:2D3D:fixed_m}).  
For non-integer dimensions the method of 
\cite{Novotny1993,Novotny2002} can be used 
to obtain the hopping terms in a translationally 
invariant manner, and the Eq.~(\ref{Eq:Proof:03}) 
can be utilized to put in the on site energies 
that satisfy the quantum dragon condition 
of Eq.~(\ref{Eq:Proof:02}) for each slice of the nanodevice.  
}
Note each hopping terms in ${\bf h}_j$ may all be chosen 
to be an independent random variate 
from identical or from 
different random distributions, while the on site energy 
terms $\epsilon_{i,j}$ are given by 
Eq.~(\ref{Eq:Proof:03}).
~~~~~~~~~{\it Q.E.D.}

 \bibliographystyle{elsarticle-num} 
 \bibliography{CSaF_Dragon,CSaF_Dragon_added}





\end{document}